\documentclass[useAMS,usenatbib]{mn2e}

\usepackage{epsfig}
\usepackage{color}

\def\cgs{$\rm ergs \, s^{-1} cm^{-2}$}

\title[X-ray vs. IR selection of distant galaxy
  clusters]{X-ray versus infrared selection of distant galaxy
  clusters: A case study using the XMM-LSS and SpARCS cluster samples}
\author[J. P. Willis et al.]{J. P. Willis$^{1}$\thanks{E-mail:
    jwillis@uvic.ca (JPW)}, M. E. Ramos-Ceja$^{2}$, A. Muzzin$^{3}$,
  F. Pacaud$^{2}$, H. K. C. Yee$^{4}$,
  G. Wilson$^{5}$\\ $^{1}$Department of Physics and Astronomy,
  University of Victoria, 3800 Finnerty Road, Victoria, BC, V8P 5C2,
  Canada\\ $^{2}$Argelander-Institut f{\" u}r Astronomie (AIfA),
  Universit{\" a}t Bonn, Auf dem H{\" u}gel 71. D-53121 Bonn,
  Germany\\ $^{3}$Department of Physics and Astronomy, York
  University, 4700 Keele Street, Toronto, Ontario, M3J
  1P3\\ $^{4}$Department of Astronomy and Astrophysics, University of
  Toronto, 50 St. George Street, Toronto, Ontario M5S 3H4,
  Canada\\ $^{5}$Department of Physics and Astronomy, University of
  California-Riverside, 900 University Avenue, Riverside, CA 92521,
  USA}
\begin{document}

\date{Accepted 2018 April 13. Received 2018 April 13; in original form 2017 June 2}

\pagerange{\pageref{firstpage}--\pageref{lastpage}} \pubyear{2002}

\maketitle

\label{firstpage}

\begin{abstract}

We present a comparison of two samples of $z>0.8$ galaxy clusters
selected using different wavelength-dependent techniques and examine
the physical differences between them.  We consider 18 clusters from
the X-ray selected XMM-LSS distant cluster survey and 92 clusters from
the optical-MIR selected SpARCS cluster survey.  Both samples are
selected from the same approximately 9 square degree sky area and we
examine them using common XMM-Newton, Spitzer-SWIRE and CFHT Legacy
Survey data.  Clusters from each sample are compared employing
aperture measures of X-ray and MIR emission.  We divide the SpARCS
distant cluster sample into three sub-samples: a) X-ray bright, b)
X-ray faint, MIR bright, and c) X-ray faint, MIR faint clusters.  We
determine that X-ray and MIR selected clusters display very similar
surface brightness distributions of galaxy MIR light.  In addition,
the average location and amplitude of the galaxy red sequence as
measured from stacked colour histograms is very similar in the X-ray
and MIR-selected samples.  The sub-sample of X-ray faint, MIR bright
clusters displays a distribution of BCG-barycentre position offsets
which extends to higher values than all other samples.  This
observation indicates that such clusters may exist in a more disturbed
state compared to the majority of the distant cluster population
sampled by XMM-LSS and SpARCS.  This conclusion is supported by
stacked X-ray images for the X-ray faint, MIR bright cluster
sub-sample that display weak, centrally-concentrated X-ray emission,
consistent with a population of growing clusters accreting from an
extended envelope of material.

\end{abstract}


\section{Introduction}

A galaxy cluster is a massive physical structure dominated by a dark
matter halo, an intra-cluster medium (ICM) consisting of a hot
atmosphere of baryonic gas, and a population of member galaxies.
Furthermore, each of the above mass components is in approximate
virial equilibrium with the total gravitational potential.

Galaxy clusters represent the most massive structures in the Universe
to have achieved this state at the present epoch \---\ with the most
extreme examples of such clusters presenting virial masses in excess
of $10^{15} \rm M_\odot$ \citep{mcdonald2012,menanteau2012}.  Defining
a lower mass limit for a galaxy cluster is more problematic.  The
physical properties of galaxy clusters, such as mass, X-ray
luminosity, X-ray temperature, optical richness or velocity
dispersion, can be related to each other via simple power laws
\citep{kaiser1986}.  Such power law relations appear to be scale free
\citep[e.g.][]{mahdavi2001,anderson2015}, an observation that makes
the definition of a minimum galaxy cluster mass an arbitrary
statement.  However, for the sake of argument, this minimum mass is
generally taken to be in the range $\log M/M_\odot = 13.5-14$
\citep[e.g.][]{sarazin1986}.

Galaxy clusters are identified by various observational techniques:
overdensity searches identify the statistical excess of projected
cluster member galaxies in relation to the ``background'' of
non-cluster galaxies along the line-of-sight
\citep[e.g.][]{postman1996,gladders2000,rykoff2014}; X-ray searches
identify the integrated emission from optically thin bremsstrahlung
emission arising from the hot ICM
\citep[e.g.][]{gioia1990,bohringer2001,clerc2012}; Weak lensing
searches identify the integrated shear signal introduced into the
shapes of background galaxies by the effect of the cluster mass on
local spacetime \citep[e.g.][]{miyazaki2002,wittman2006,gavazzi2007};
and Sunyaev-Zeld'ovich (SZ) searches identify the apparent decrement
in the brightness of the cosmic microwave background (CMB) caused by
the inverse Compton scattering of CMB photons by energetic electrons
in the ICM \citep[e.g.][]{stan2009,marriage2011,reichardt2013}.

Each of these observational techniques is sensitive to a distinct
physical component of galaxy clusters: Overdensity searches are
sensitive to the integrated star formation history of the cluster;
X-ray searches to the projection of the square of the free electron
density in the ICM (with a weak dependence upon the ICM temperature);
Weak lensing searches to the total projected cluster mass; and SZ
searches to the thermal pressure of the free ICM electrons (with small
relativistic corrections for the hottest systems).

The question which concerns this paper is how do the physical
properties of cluster samples differ depending upon the observational
technique used to identify them?  Such knowledge is important because
it a) permits a consistent comparison between results generated for
different cluster samples
\citep[e.g.][]{gilbank2004,barkhouse2006,rozo2014b,rossetti2017} and
b) provides a means to relate the results generated for a particular
cluster sample to the ``true'' cluster population
\citep[e.g.][]{borgani2001b,mantz2008,rozo2010,planck2014,jimeno2017}.

At redshifts $z<0.3$ the comparison between cluster catalogues
compiled using different wavebands is mature.  \cite{rozo2014a} compare
three such low redshift cluster catalogues, selected using optical,
X-ray and SZ techniques and demonstrate that each sample can be
projected onto a statistically consistent set of common scaling
relations.

An alternative approach is to compare cluster catalogues compiled over
common sky areas on the basis of individual detections as has been
performed by \cite{berge2008} using X-ray and weak lensing selected
clusters and \cite{starikova2014} using X-ray, weak lensing and
spectroscopically identified clusters.  Although to date such
comparisons have been limited to small sky areas (4 square degrees or
less) and consequently small sample sizes, the conclusions are that
clusters that are detected in one waveband but are absent in another
can generally be explained by measurement uncertainty and scatter in
the relationships between observables and mass.

Two studies which attempt to account for unmatched as well as matched
clusters between catalogues have been conducted by \cite{donahue2002}
and \cite{sadibekova2014}, respectively. Each compares an optical and
X-ray selected cluster sample typically sensitive to clusters at
$z<1$.  Once again, Donahue et al. noted that the relative fraction of
optical clusters lacking an X-ray counterpart could be explained by a
steep scaling relationship between X-ray and optical cluster
luminosity, i.e. $L_X \propto L_{opt}^{3-4}$. In addition, both
studies concluded that the majority of the X-ray clusters lacking an
optical counterpart could be attributed to the maximum and minimum
effective redshift limitations of the optical cluster selection
criteria.

This paper is motivated by the interest in applying a similar
comparison to samples of distant galaxy clusters, in this case
compiled using data at X-ray and mid-infrared (MIR) wavelengths.  This
motivation is in part due to the increased potential for cluster-scale
astrophysics to influence the observed properties of such clusters and
their galaxy populations, e.g. recent star formation
\citep{bayliss2014,hayashi2011,brodwin2013,nantais2016,nantais2017},
mergers \citep{nastasi2011,lotz2013}, or AGN activity
\citep{galametz2009,martini2013,ehlert2015,alberts2016}.  However,
because such distant clusters are typically identified at low
significance in survey quality data (and thus might be prone to
considerable scattering effects on mass-observable relations) we do
not compare cluster catalogues compiled at different wavebands via
their scaling relations.  Instead we apply multiple techniques to
determine the extent to which each sample exhibits different physical
properties and relate those properties to cluster evolutionary state
(e.g. star formation and merger histories).

The structure of the paper is as follows: In Section 2 we describe
each distant cluster sample. In Section 3 we describe the available
data common to each sample and in Section 4 we use this data to
measure fixed aperture brightness values of individual clusters. We
present the results of this analysis in Section 5 and draw appropriate
conclusions in Section 6. Where necessary we assume a
Friedmann-Lem{\^a}itre-Robertson-Walker cosmological model described
by the parameters $\Omega_M = 0.3$, $\Omega_\Lambda=0.7$, $H_0=70 \rm
kms^{-1}Mpc^{-1}$.

\section{Distant galaxy cluster samples}

\subsection{XMM-LSS}

The distant X-ray selected galaxy clusters studied in this paper are
taken from \cite{willis2013}.  This paper presented a sample of 20
galaxy clusters at $z>0.8$ selected from an approximately 9 square
degree area of the XMM-LSS survey \citep{pierre2004}. Extended X-ray
sources were classified as C1 if they satisfied the criteria {$\tt
  detection likelihood > 32, extension likelihood > 33, extension >
  5\arcsec$} and as C2 if they satisfied {$\tt extension likelihood >
  15, extension > 5\arcsec$} (see \citealt{pacaud2006} for further
details).  In addition, sources were classified as distant clusters if
they either possess a known spectroscopic redshift $z>0.8$ or display
a line of sight overdensity of galaxies unlikely to be located at
$z<0.8$. Ten band photometry ($ugrizYJK[3.6\micron][4.5\micron]$) for
these latter systems was employed to derive photometric redshifts for
bright galaxies deemed to be associated with each extended X-ray
source.  Candidate clusters were retained in the distant sample if
they displayed an overdensity of galaxies consistent with a single
location in photometric redshift space at $z_{phot}>0.8$. Of these 20
clusters, 18 lie within the common footprint of the
XMM-LSS/SWIRE/CFHTLS-W1 surveys and were retained for analysis.

\subsection{Spitzer Adaptation of the Red Sequence Cluster Survey} 

The MIR selected distant clusters studied in this paper are taken from
the Spitzer Adaptation of the Red Sequence Cluster survey
\citep[SpARCS;][]{muzzin2009,wilson2009}.  Candidate clusters are
identified with significant overdensities in the multiple dimensional
space defined by sky position, $z^\prime-3.6\micron$ colour, and
3.6\micron\ brightness.  Two centroid estimates are provided for each
cluster: the first is the sky position of the brightest cluster galaxy
(BCG) determined from its location on the cluster red sequence. The
second estimate is referred to as the barycentre position and is based
upon the mean sky position of all candidate cluster members identified
by the red sequence method.  We employ both centroid measures in this
paper and comment explicitly on how each produces different results.
Finally, in addition to the candidate cluster sky position, we also
retain the cluster photometric redshift estimate based upon the
location of the cluster red sequence in $z^\prime-3.6\micron$ colour.

The SpARCS catalogue located within SWIRE field contains 218 candidate
galaxy clusters within the redshift interval $0.1<z<1.7$.  These
correspond to clusters satisfying a richness cut of $N_{red} > 6$,
where $N_{red}$ is the number of background-subtracted red-sequence
galaxies brighter than $M^\ast(z)+1$. The value of $M^\ast(z)$ is
computed from a passive stellar population evolution model formed at
$z_f=4$ \citep{muzzin2008}.  Red-sequence galaxies are defined as
those within $\pm 0.15$ magnitudes of the best-fitting $z^\prime-3.6$
model for each cluster.  The threshold $N_{red}>6$ was set in order to
create a sample of clusters suitable for comparison that balanced
completeness and purity.  From this sample we selected those clusters
located within the the combined XMM-LSS/SWIRE/CFHTLS-W1 footprint and
lying at $z>0.8$ for comparison with XMM-LSS.

Although SpARCS employs a richness-based estimator to identify cluster
candidates, the inclusion of colour and luminosity information limits
the contamination rate arising from projected large-scale structure to
5\%\ at $z<0.6$ \citep{gladders2000,gilbank2018} and to $\sim
15$\%\ at $z\sim 1$ \citep{gilbank2018}.  In addition, of the 10
SpARCS clusters observed as part of the Gemini Cluster Astrophysics
Spectroscopic Survey (GCLASS; \citealt{muzzin2012}), only 1 system,
J104737$+$574137, shows limited evidence of line-of-sight galaxies
close to but not at the redshift of the cluster being erroneously
assigned cluster membership. The potential also exists for distant
clusters to generate multiple detections, especially where each
detection is composed of a small number of faint galaxies.  With only
three pairs of $z>0.8$ clusters identified within 1\arcmin\ of each
other on the sky this is not a significant issue.  However, in these
cases we applied a selection to the SpARCS catalogue to accept the
cluster generating the higher signal-to-noise ratio (SNR) detection
along the line of sight.  These considerations resulted in a sample of
92 clusters for analysis.

\subsection{Catalogue matching}

An initial means of assessing the commonality of objects appearing in
each sample is to match the catalogue positions using a fixed
tolerance \citep[e.g.][]{oguri2017}.  We performed such a test using a
tolerance of 60{\arcsec} (having confirmed that this arbitrary
threshold does not exclude clusters close to, yet exceeding, this
value).
\begin{table*}
\caption{Matching results between the XMM-LSS and SpARCS $z>0.8$
  cluster samples. The column ``ID'' refers to the cluster identifier
  presented in Willis et al. (2013) and the column ``XLSSC'' lists the
  short XMM-LSS identifier. The matching code B indicates a cluster
  match using the BCG centroid and C indicates a match using the
  barycentre position. The identifier NM indicates that no match was
  obtained within the 60\arcsec\ tolerance.  Where $N_{red} \leq 6$,
  this indicates a match obtained with a cluster lying below the
  richness cut applied to the SpARCS sample used for analysis in this
  paper. The X-ray mass values correspond to $M_{200c}$ are taken from
  Willis et al. (2013) and are computed by applying appropriate
  scaling relations to the measured X-ray flux and redshift. As
  discussed in Willis et al. (2013) such mass estimates vary
  considerably with the assumed scaling relation model and should be
  take as indicative, as opposed to exact, values.}
\label{matching_table}
\centering
\begin{tabular}{clcccccccc}
\hline
ID & Cluster name                          & XLSSC & Class & Spec. & XMM-LSS & SpARCS & Match & $N_{red}$ & X-ray mass \\
&&&&confirmed&redshift&$z_{phot}$&&& ($\times 10^{14} \rm M_\odot$) \\     
\hline
01 &XLSS\hspace{2.5mm}  J022400.4-032529   &  32 & C2 & Y               & 0.803                   & 0.98         & BC & 11.5 & $1.21_{-0.14}^{+0.25}$ \\
02 &XLSS\hspace{2.5mm}  J022233.8-045803   &  66 & C2 & Y               & 0.833                   & 0.92         & BC &  8.5 & $1.09_{-0.18}^{+0.07}$ \\
03 &XLSSU J021832.0-050105                 &  64 & C2 & Y               & 0.875                   & 0.98         & BC & 16.5 & $1.26_{-0.08}^{+0.08}$ \\
04 &XLSSU J021524.1-034332                 &  67 & C1 & Y               & 1.003                   & 1.08         & BC & 15.8 & $2.96_{-0.19}^{+0.21}$ \\
05 &XLSS\hspace{2.5mm} J022253.6-032828    &  48 & C1 & Y               & 1.005                   & 1.13         & BC &  6.1 & $1.45_{-0.17}^{+0.19}$ \\
07 &XLSS\hspace{2.5mm}  J022404.1-041330   &  29 & C1 & Y               & 1.050                   & 1.10         & BC & 11.9 & $2.35_{-0.15}^{+0.15}$ \\
08 &XLSS\hspace{2.5mm}  J022709.2-041800   &   5 & C1 & Y               & 1.053                   & 1.18$^\ast$  & BC & 10.3 & $1.17_{-0.13}^{+0.07}$ \\
09 &XLSS\hspace{2.5mm}  J022303.3-043621   &  46 & C2 & Y               & 1.213                   & 1.4$^\ast$   & BC & 9.2  & $0.92_{-0.15}^{+0.11}$ \\
12 &XLSSU J021547.7-045027                 &  78 & C1 & Y               & 0.953                   & 1.06         & BC & 4.8 & $1.18_{-0.14}^{+0.07}$ \\
13 &XLSSU J021859.5-034608                 &     & C2 & Y               & 0.979                   & 1.06         & BC & 5.8 & $1.20_{-0.14}^{+0.16}$ \\
14 &XLSS\hspace{2.5mm}  J022059.0-043921   &     & C2 &                 & $1.11^{+0.29}_{-0.26}$    & 1.23         & C  & 4.5  & $1.06_{-0.12}^{+0.14}$ \\
15 &XLSS\hspace{2.5mm}  J022252.3-041647   &     & C2 &                 & $1.12^{+0.18}_{-0.17}$    & N/A          & NM &    --& $0.84_{-0.09}^{+0.10}$ \\
16 &XLSSU J021712.1-041059                 &     & C2 &                 & $1.48^{+0.25}_{-0.10}$    & N/A          & NM &    --& $0.97_{-0.20}^{+0.12}$ \\
17 &XLSSU J021700.3-034747                 &     & C2 &                 & $1.54^{+0.30}_{-0.31}$    & N/A          & NM &    --& $0.87_{-0.14}^{+0.16}$ \\
18 &XLSSU J022005.5-050824                 &     & C2 &                 & $1.65^{+0.25}_{-0.26}$    & N/A          & NM &    --& $1.17_{-0.19}^{+0.15}$ \\
20 &XLSS\hspace{2.5mm}  J022418.7-043959   &     & C2 &                 & $1.67^{+0.20}_{-0.20}$    & 1.63         & BC & 4.8  & $0.78_{-0.28}^{+0.26}$ \\
21 &XLSSU J021744.1-034536                 & 122 & C1 & Y               & 1.98                    & N/A          & NM &    --& $1.33_{-0.15}^{+0.17}$ \\ 
22 &XLSS\hspace{2.5mm}  J022554.5-045058   &     & C2 &                 & $2.24^{+0.26}_{-0.24}$    & N/A          & NM &    --& $0.59_{-0.09}^{+0.07}$ \\
\hline 
\multicolumn{10}{l}{Note: In the two cases marked by an asterisk
  each has a $z\sim 1.8$ SpARCS cluster providing the closest spatial
  match;}\\ 
\multicolumn{10}{l}{however, the selected clusters are both
  within the 60\arcsec\ matching radius and at similar redshifts.}
\end{tabular}
\end{table*}
The results of the matching procedure are summarised in Table
\ref{matching_table} and indicate that 11/12 XMM-LSS clusters at
$0.8<z<1.4$ are matched to a SpARCS $z>0.8$ cluster, while at
$1.4<z<1.7$ the fraction is 1/4. A further two XMM-LSS clusters (IDs
21 and 22) lie beyond the maximum redshift included in the SpARCS red
sequence catalogue ($z=1.7$).

Although the matching fraction for the highest redshift XMM-LSS
clusters is low, we note that, in addition to their very high
redshift, they possess relatively low masses and display varying red
sequence populations (Figure 12 of \citealt{willis2013}). It therefore
remains plausible that the unmatched, high-redshift XMM-LSS clusters
are real clusters that exhibit low stellar mass red sequences such
that they do not pass the SpARCS selection criteria.

We performed a second matching analysis, this time comparing SpARCS
$z>0.8$ cluster positions to all detected X-ray sources in the common
footprint of the two samples, irrespective of their redshift or
morphological classification (i.e. C1, C2, C3 or point source; as
described in \citealt{pacaud2006}).  In this case 33/92 SpARCS $z>0.8$
clusters were matched within 30{\arcsec} to a source in the full
XMM-LSS catalogue, rising to 50/92 within a 60{\arcsec} matching
radius. The majority of X-ray sources contributing to these matches
are low-SNR sources of marginally significant extent (C3) in addition
to low-SNR point sources (which include unresolved extended
sources). However, if we restrict the matching analysis solely to 12
X-ray bright SpARCS clusters (see Section \ref{sec_flux_flux} for
details of X-ray aperture flux measurement), where one would expect a
match to the X-ray source catalogue to occur, we find 6/12 and 11/12
matches within 30 and 60{\arcsec} respectively. Of these, 3 and 4
respectively are associated with C1 or C2 sources.

Relaxing the restriction on morphological type of X-ray source to
which the SpARCS clusters are matched generates a larger fraction of
matched objects.  This indicates that X-ray sources classified as C1
or C2 within the XMM-LSS survey form a restricted subset of a larger
population of gravitationally-bound structures as traced by the SpARCS
$z>0.8$ cluster sample.  Furthermore, as will be shown in Section
\ref{sec_flux_flux}, the majority of SpARCS $z>0.8$ clusters display
X-ray aperture fluxes which fall significantly below those determined
for distant XMM-LSS clusters with the result that the XMM-LSS source
detection pipeline fails to recognise them as individual X-ray
sources.  This is not surprising as the XMM-LSS C1/C2 classification
is designed to identify highly significant, extended sources as galaxy
clusters in order to construct samples suitable for precision
cosmological analyses (following correction for the survey selection
function to account for the partial view of the total cluster
population).

What is clear however, is that performing a matching analysis will
only provide a limited understanding of the physical differences
between the XMM-LSS and SpARCS cluster samples. The subsequent
analyses in this paper therefore consider the measured properties of
individual clusters in each sample at both X-ray and MIR wavelengths
as a more direct means of determining the extent and nature of bulk
physical differences between the two cluster samples.

\section{Data}

\subsection{X-ray data}

X-ray data are obtained from the XMM-LSS survey.  The survey has
imaged a 11.1 square degree area centered on R.A. $=2^h 22^m$,
Dec. $=-$4\degr30\arcmin\ with a mosaic of 93 overlapping XMM-Newton
pointings \citep{chiappetti2013}.  Each pointing displays a typical
exposure time of 10 ks and corresponds to a single observation with
the EPIC detectors (MOS1, MOS2 and PN) in full frame imaging mode,
spanning a field of view of roughly 30\arcmin\ diameter.  The
effective flux limit for extended sources identified by the C1/C2
surface brightness selection threshold is $\sim 1 \times 10^{-14} \,
\rm ergs \, s^{-1} cm^{-2}$.

\subsection{Spitzer MIR data}

Approximately 9 square degrees of the XMM-LSS region has been imaged
by the Spitzer space telescope as part of the SWIRE extragalactic
survey \citep{lonsdale2003}.  The Spitzer/SWIRE data used in this
paper are described in \cite{chiappetti2013}.  In particular, we make
use of the IRAC channel 1 data corresponding to a photometric bandpass
located at 3.6\micron. Fluxes for extended sources were measured
within the so-called aperture 2\footnote{This aperture corresponds to
  a 1\farcs9 radius circle with an applied aperture correction which
  approximates to a total brightness measure. See Section 4.11.2 of
  the IRAC instrument handbook or {\tt
    http://irsa.ipac.caltech.edu/data/SPITZER/docs/irac/iracinstrumenthandbook/29/}}
and are expressed in AB magnitudes employing the relation $\rm
[3.6]=23.9 - 2.5 \log (f_\nu/1 \mu Jy)$.

\subsection{Optical data}

\begin{figure}
\psfig{figure=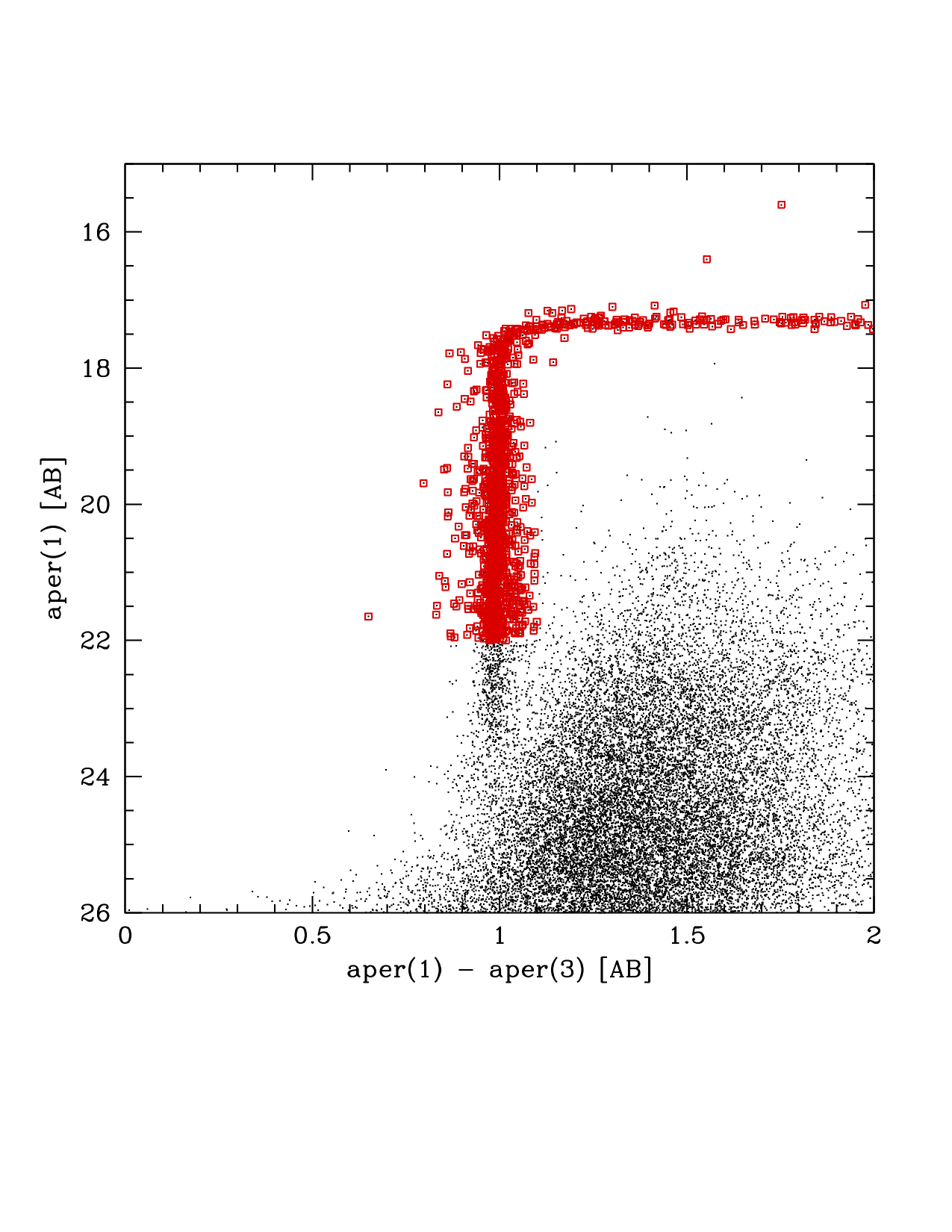,width=3.0in,angle=0.0}
\caption{An example of star/galaxy separation applied to a particular
  $r^\prime$-band image of the CFHTLS data set (``Field A''). The red
  points indicate sources identified as stars. Star-galaxy separation
  is performed for sources satisfying {\tt aper(1)} $<$ 22 AB. No
  separation is performed fainter than this threshold.}
\label{fig_stargal}
\end{figure}

In addition to the X-ray and MIR data described above we also used
optical $u^\ast g^\prime r^\prime i^\prime z^\prime$ photometry
obtained from the Canada France Hawaii Telescope Legacy Survey W1
field \citep[CFHTLS-W1;][]{gwyn2012}. Photometry was computed within
an aperture based upon the \cite{kron1980} radius and quoted on the AB
magnitude system.  Optical photometry was identified for all sources
within the Spitzer MIR catalogue described above. Two independent
multi-band catalogues were produced using {\tt SExtractor} in
two-image mode with either the $r^\prime$- or $z^\prime$-band image
used as the detection band in each case.  The construction of separate
$r^\prime$- and $z^\prime$-selected later permitted a self-consistent
examination of the effects of the colour selection of high-redshift
cluster galaxies using either $r^\prime-3.6 \micron$ or $z^\prime-3.6
\micron$ colours.  Optical sources were matched to MIR sources when
they are located within a positional tolerance of $<2\arcsec$ with the
brightest optical source selected in the case of multiple matches.  In
the case where a matching optical source was not found, the computed
colour of the Spitzer source represents a lower limit based upon the
completeness of the optical catalogue in the appropriate band.
Finally, star-galaxy separation was performed upon the matched
catalogue in the appropriate optical detection band using the
distribution of sources on the plane defined by {\tt aper(1)-aper(3)}
versus {\tt aper(1)} where {\tt aper($n$)} is the magnitude measured
in an aperture of diameter equal to $n$ arcseconds (these quantities
represent measures of source extent and brightness respectively;
Figure \ref{fig_stargal}).

\section{Fixed aperture brightness measures}

Given the two cluster samples described above and a common X-ray,
optical and MIR data set, it is possible to measure X-ray and MIR
brightness measures for all clusters employing a simple, consistent
approach.

The cluster signal in each waveband was measured in a circular
aperture of fixed radius equal to 1\arcmin\ centered on the X-ray
position for XMM-LSS clusters and either the BCG or the barycentre
position for SpARCS clusters.  This approach was selected to measure
the cluster signal in as robust a manner as possible and using the
fewest assumptions regarding the properties of individual
clusters. For example, this approach requires only the sky position of
each cluster and thus lends itself well to comparing cluster samples
drawn from a variety of selection approaches.

Application of a circular aperture is the simplest response to the
lack of data on the shapes of distant clusters.  Furthermore,
application of a fixed angular radius offers a number of advantages:
the background applied to correct the line-of-sight signal from each
cluster is uniform across the sample. As the line-of-sight signal from
each cluster is often background dominated, this generates consistent
and comparable uncertainties across the sample of measurements.

In addition, although one could choose to apply an aperture of fixed
physical radius in the rest frame of each cluster it should be noted
that, over the redshift interval $0.8<z<2$ and within the assumed
cosmological model, the angular diameter distance (required to convert
from angular to physical radius) varies by approximately $\pm5\%$
about the fiducial redshift $z=1$.

\subsection{X-ray aperture photometry}

The X-ray brightness measurement from a galaxy cluster is sensitive to
the emission from gravitationally confined gas at the virial
temperature of the cluster gravitational potential. The X-ray
brightness of a cluster is primarily a measure of the (square of) the
baryonic gas mass with a slowly varying dependence upon the cluster
gas temperature.

We implemented a Bayesian approach for calculating X-ray aperture
photometry. The method is an adaptation from the approach taken by
\cite{vanDyk01} and \cite{park2006}. In the following we explain our
procedure.

Net source counts are computed from independent source and background
apertures areas.  Sky areas associated with non-extended source
detections (i.e. non-C1 or C2) were masked from this process to remove
any possible contribution from X-ray Active Galactic Nuclei (AGN).  We
assume that $C$ counts are measured in a source aperture of area
$A_s$, and $B$ counts are measured in a background aperture of area
$A_b$. The observed counts are generated via a Poisson process, i.e.,
\begin{eqnarray}
 C&\sim& \textrm{Poisson}(f(s+b)),\\
 B&\sim& \textrm{Poisson}(rgb),
\end{eqnarray}
where $C$ is given by the sum of counts due to the source, $s$, and
the background, $b$. The symbols $f=1/T_{s}$ and $g=1/T_{b}$ are
factors that convert the net counts to count-rates given the average
exposure time in $A_{s}$ and $A_{b}$. The quantity $r$ is an area
correction factor given by $r=A_{b}/A_{s}$. We determine $s$ from its
posterior probability density marginalized over the background,
\begin{equation}
 p(s|CB)=\int \mathrm{d}b~p(sb|CB).
\label{pp_s}
\end{equation}
Via Bayes' theorem, the joint posterior probability of $s$ and $b$ can
be rewritten as
\begin{equation}
 p(sb|CB)=\frac{p(s)p(b)p(C|sb)p(B|b)}{\int \int \mathrm{d}b \mathrm{d}s~p(s)p(b)p(C|sb)p(B|b)}.
\end{equation}
$p(C|sb)$ and $p(B|b)$ are Poisson distributions, and $p(s)$ and
$p(b)$ are generalized $\gamma$-priors. Further details in the final
analytical derivation of $p(s|CB)$ can be found in Appendix A2 of
\citet{park2006}.  In our work we assume non-informative priors, given
as the result
\begin{equation}
{
p(s|CB) = A^{-1} \times B,
}
\label{pp_s_final}
\end{equation}
where
\begin{equation}
{
A = \Bigg[\sum^C_{j=0}\frac{1}{\Gamma(j+1)\Gamma(C-j+1)}\frac{\Gamma(C+B+1-j)}{(f+gr)^{C+B+1-j}}\frac{\Gamma(1+j)}{f^{1+j}}\Bigg]
}
\end{equation}
and
\begin{equation}
{
B = \sum^C_{j=0}\frac{1}{\Gamma(j+1)\Gamma(C-j+1)}\frac{\Gamma(C+B+1-j)}{(f+gr)^{C+B+1-j}}s^{j}e^{-fs}.
}
\end{equation}
The value of $s$ is obtained from the mode of $p(s|CB)$ distribution,
and the confidence levels are determined by numerically integrating
$p(s|CB)$ until the desired confidence level is reached. In this way,
if the mode of $p(s|CB)$ is equal to 0, one can still provide upper
limit for $s$.

We applied the above approach to the clusters in the sample to obtain
their X-ray aperture photometry. Apertures of 1\arcmin\ radius are
used and are located on the cluster centroid. We have two ways to
obtain the background counts depending on the cluster position in the
XMM pointing:
\begin{itemize}
 \item If the cluster centroid is close ($<$2\arcmin) to the pointing
   center, the background aperture is defined as an annulus centered
   on the cluster position. The annulus has a width of 1\arcmin, and
   is 1\arcmin\ away from the source aperture to avoid contamination
   from the cluster.
 \item If the cluster centroid is far ($>$2\arcmin) from the pointing
   center, the background aperture is an annulus encompassing the
   cluster aperture and at similar off-axis angle. A quadrant of
   $45$~degrees (centered on the cluster) is excluded from the
   background measurement to avoid residual cluster contamination.
\end{itemize}
This approach accounts for the radial variation of the XMM
background. The procedure is applied separately to each XXM-Newton
EPIC detector, obtaining three different count-rate posterior
probability density distributions for each cluster. We then convert
each count-rate distribution into a flux posterior probability
density, $p(f)$, through an energy conversion factor (ECF). This
factor is calculated using {\tt XSPEC} \citep{arnaud1996} and an {\tt
  APEC} emission model with $z=1$, $T=2$~keV,
$N_H=2.6\times10^{20}~$cm$^2$, $Ab=0.3$, and standard on-axis EPIC
response matrices. The ECF scales the flux of the three EPIC detectors
to a common sensitivity. The final flux posterior probability density,
$p(f_X)$, for a given cluster is obtained by multiplying the
individual flux distributions of the different EPIC detectors:
\begin{equation}
{
 p(f_X)=\prod^{3}_{i=1} p(f_{i}),
}
\end{equation}
where $i$ refers to the three EPIC detectors. The final X-ray flux is
obtained from the mode of $p(f_X)$, together with its corresponding
$68\%$ confidence levels.

\subsection{Spitzer MIR aperture photometry}
\label{spitzer_apphot}

The Spitzer MIR brightness measurement for a galaxy cluster is
computed as the summed stellar 3.6\micron\ brightness of individual
galaxies identified as cluster members.  To determine membership we
applied a colour cut to the optical-MIR source catalogue to identify
candidate $z>0.8$ galaxies \citep[e.g. ][]{muzzin2008}.  

We investigated the application of two cuts designed to select $z>0.8$
passive galaxies, namely $r^\prime-3.6\micron>3.4$ and
$z^\prime-3.6\micron>1.3$.  The MIR fluxes of those galaxies that
satisfy these colour cuts and lie within 1\arcmin\ of the cluster
centroid are then summed to provide a two separate $3.6\micron$ flux
measurements per cluster.  Much of the analysis in the subsequent
sections was repeated using aperture fluxes and luminosities computing
using either the $r^\prime-3.6\micron$ and $z^\prime-3.6\micron$ cuts.
However, at this point we note that all of the analysis which follows
generated similar results and conclusions irrespective of which colour
cut was considered.  The use of two colour cuts therefore provided
useful a consistency check but, in the interests of brevity, we only
present results derived using the $r^\prime-3.6\micron$ colour cut.

Each aperture measurement was corrected for unassociated galaxies
along the line of sight employing a sample of 5000 randomly placed
1\arcmin\ apertures located within the common survey footprint and
using the same photometric thresholds as applied to the cluster
samples.  Although most non-cluster galaxies will lie in the
foreground of a distant cluster, for simplicity we refer to these
randomly placed apertures as ``the background''.  We rejected
background apertures which lay within 2\arcmin\ of any SpARCS cluster
to retain a sample of 4667 apertures for analysis.  The distribution
of background aperture flux measurements was modelled as a Gaussian
function modified by a shallow power law to describe a slight skewness
of the observed distribution toward higher background values.  The
source flux in each cluster aperture was then computed as the maximum
value of the posterior distribution of the cluster aperture flux
(source plus background) minus the background model with the prior
that the source flux $s \ge 0$.  The error on each cluster flux
measurement is computed from the interval of the above posterior
distribution per cluster which contains 68\%\ of the distribution.

\section{Results}

\subsection{X-ray versus MIR flux and luminosity measures}
\label{sec_flux_flux}

\begin{figure*}
\centering
\psfig{figure=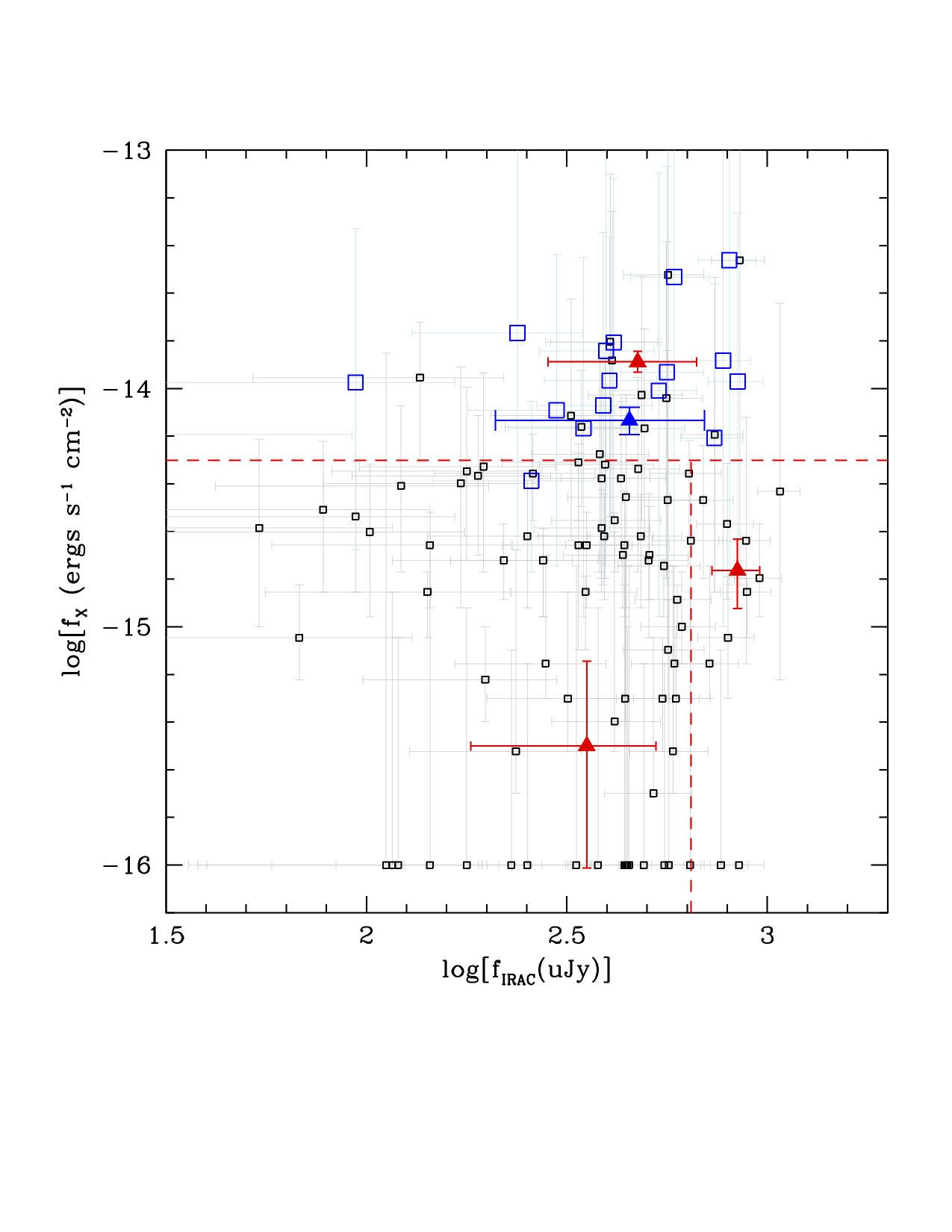,width=3.0in,angle=0.0}
\psfig{figure=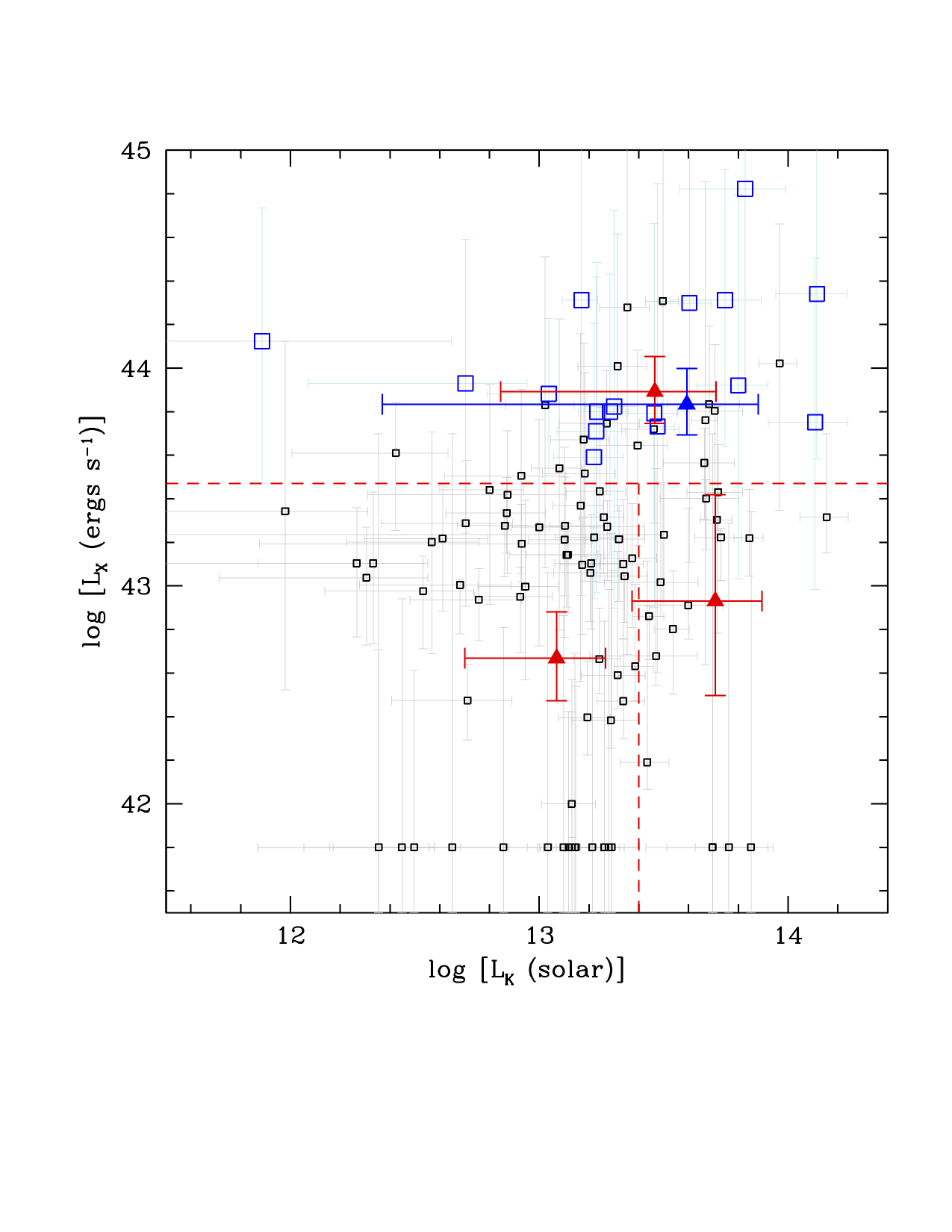,width=3.0in,angle=0.0}
\caption{A comparison of X-ray and MIR 1\arcmin\ aperture brightness
  values for XMM-LSS (blue squares) and SpARCS (black squares)
  clusters. This comparison uses the $r-3.6\micron>3.4$ colour cut and
  employs the BCG position of each SpARCS cluster. Error bars indicate
  the 68\%\ interval of the posterior background subtracted flux
  distribution for each source. The red dashed lines indicate the
  selection cuts applied to generate sub-samples of the SpARCS cluster
  for further analysis. Left: Comparison by flux. SpARCS clusters with
  zero measured X-ray flux are marked at $\log f_X = -16$ for clarity.
  The solid triangles indicate the fluxes measured for the stacked
  cluster samples (see Section \ref{xstack}); red indicates SpARCS and
  blue indicates XMM-LSS clusters. Right: Comparison by
  luminosity. SpARCS clusters with zero measured X-ray luminosity are
  marked at $\log L_X = -41.8$ for clarity. See text for further
  details.}
\label{fig_xflux}
\end{figure*}

Figure \ref{fig_xflux} compares the cluster 3.6\micron\ summed
aperture flux to the X-ray aperture flux measured for 18 XMM-LSS and
92 SpARCS $z>0.8$ clusters. This comparison uses the
$r^\prime-3.6\micron>3.4$ colour cut and employs the BCG position of
each SpARCS cluster.

Two initial impressions are apparent from this comparison.  Firstly,
there is a broad correlation in luminosity defined by X-ray faint, MIR
faint ranging to X-ray bright, MIR bright clusters.  We do not attempt
to quantify this trend in the current study as the measurement
approach taken is deliberately simple and is designed to provide a
robust comparison between clusters of widely different properties.
Secondly, in contrast to this broadly defined correlation between
X-ray gas emission and stellar emission within both cluster samples,
there exist a number of MIR selected clusters which, although they are
among the brightest MIR sources in either sample, appear to be
relatively deficient in measured X-ray aperture flux or luminosity.
Some caution is required: there is much scatter in the measured
distributions displayed in Figure \ref{fig_xflux} and the apparently
MIR bright, X-ray faint clusters mentioned above do not represent
significant outliers.  However, the distributions displayed in Figure
\ref{fig_xflux} are sufficiently interesting to investigate whether
splitting the SpARCS clusters into sub-samples based upon their
measured aperture fluxes identifies physically distinct clusters.

In the following sections we compare the physical properties of the
XMM-LSS sample to three sub-samples based upon the distribution of the
SpARCS clusters on the X-ray/MIR plane displayed in Figure
\ref{fig_xflux} \---\ the aim being to determine if each sub-sample
displays quantifiable physical differences.  We define X-ray bright
MIR selected clusters as those displaying $f_X \ge 0.5 \times
10^{-14}$ \cgs.  We further split X-ray faint ($f_X<0.5 \times
10^{-14}$ \cgs) MIR selected clusters into those which are MIR bright
($f_{3.6\micron} \ge 650 \, \rm \mu Jy$) and faint ($f_{3.6\micron} <
650 \, \rm \mu Jy$).  These thresholds are defined arbitrarily yet
identify the broad trends present in the above diagrams, the most
important of which appears to be that a significant fraction of the
brightest SpARCSs cluster, whether defined by MIR flux or luminosity,
appear deficient in X-ray emission. Though defined in this manner it
is important to note that the conclusions presented in this paper are
relatively insensitive to the exact choice of threshold values
applied. The final numbers of clusters present in each sample are as
follows: XMM-LSS, 18; SpARCS X-ray bright, 12; SpARCS X-ray faint, MIR
bright, 10; SpARCS X-ray faint, MIR faint, 70.

There exists the concern that the X-ray faint, MIR bright sub-sample
of SpARCS clusters could be due to intrinsically X-ray faint, MIR
faint clusters boosted by high local background values that are
under-subtracted by the modal value of the global background applied
in Section \ref{spitzer_apphot}.  To investigate this issue we
computed local backgrounds about each SpARCS cluster using a circular
annulus at a fixed radial distance from the cluster centroid.
Applying a background annulus with inner and outer radii respectively
5 and 6 arcminutes from each cluster centroid generated a set of
background values distributed symmetrically about the global
background computed in Section \ref{spitzer_apphot} with no evidence
of clusters with high MIR aperture flux measurements displaying
enhanced local background values.  Identification of X-ray faint, MIR
bright clusters in Figure \ref{fig_xflux} is therefore not influenced
by the application of a global background correction to MIR cluster
aperture fluxes.

It is furthermore unlikely that the non-cluster masking procedure
applied to the computation of X-ray aperture fluxes results in the
exclusion of bona-fide cluster flux from the aperture measurement. As
mentioned previously, XMM-LSS pipeline detections corresponding to
non-C1/C2 sources were masked from the aperture measurement. An
alternative masking procedure was tested, this time excluding source
areas corresponding only to point sources classified as P1 (high
significance, low extension), with little or no qualitative difference
in the results presented above.

We also compare the X-ray and MIR brightness measures of each cluster
sample in terms of their luminosity using the measured redshift for
each cluster and the assumed cosmological model to convert the
measured aperture flux in either the X-ray or IRAC1 band respectively
to a rest-frame [0.5-2] keV X-ray and $K$-band stellar luminosity.

In addition to the luminosity distance to each source, the correction
to luminosity requires the application of a $k$-correction.  For the
X-ray brightness this correction is computed using the same $T=2$ keV
plasma model used to convert count rates to flux.  In the MIR we
employ a $k$-correction derived from a 1 Gyr, solar metallicity burst
of star formation which evolves passively from a formation redshift
$z_f=10$.This model is taken from \cite{willis2013} and matches the
red sequence evolution of the XMM-LSS distant cluster
sample. Furthermore, as demonstrated by \cite{burg2014}, the stellar
mass in SpARCS distant clusters is dominated by $M^\ast$ galaxies
located on the red sequence. The evolving spectrum was realised using
the GALAXEV 2003 stellar population synthesis code \citep{bc2003}.

Computing the MIR luminosity of a galaxy cluster within a spatial
aperture effectively involves integrating the cluster galaxy
luminosity function (LF) down to a sensitivity limit which itself may
be expected to be a function of redshift. In order to determine
whether this is likely to be an important bias in the aperture
luminosity values computed for this paper we performed the following
analysis: taking the 3.6\micron\ flux limit of the optical-MIR
catalogue as 3.7 $\mu$Jy we computed the corresponding $K$-band
luminosity of a galaxy with this flux over the redshift interval
$0.8<z<1.6$ (the extent of the SpARCS sample).  We next considered
this luminosity as the limiting value as a function of redshift to
which we integrated the $K$-band cluster galaxy LF presented by
\cite{depropris2007}. The relative change in the value of this
integral over a function of redshift indicates the expected effect on
the measured aperture luminosity values.  This LF correction factor
varies with respect to a fiducial value at $z=1$ by $\pm5\%$ over the
interval $0.8<z<1.2$ (which includes over 85\%\ of the SpARCS
clusters) and by 20\%\ at the maximal redshift $z=1.6$ of the SpARCS
catalogue. These correction factors are applied to the aperture
luminosity measurements shown in Figure \ref{fig_xflux} yet we note
that they are small compared to the measurement errors (which are
dominated by background variations).

We apply X-ray and MIR brightness cuts to the aperture luminosity
measurements in an analogous manner to those applied to the flux data.
Recall that the XMM-LSS and SpARCS samples contain 18 and 92 clusters
respectively. We define X-ray luminous MIR selected clusters as those
displaying $L_X \ge 0.3 \times 10^{44}$ \cgs.  We further split X-ray
faint ($L_X<0.3 \times 10^{44}$ \cgs) MIR selected clusters into those
which are IR luminous ($L_K \ge 2.5 \times 10^{13} \, \rm L_\odot$)
and faint ($L_K < 2.5 \times 10^{13} \, \rm L_\odot$).  The final
numbers of clusters present in each sample are as follows: XMM-LSS,
18; SpARCS X-ray bright, 17; SpARCS X-ray faint, IR bright, 16; SpARCS
X-ray faint, IR faint, 59. The is considerable overlap between cluster
sub-samples defined by flux and luminosity. For example of the 12 high
flux SpARCS clusters defined using the BCG position, 11 are present in
the luminosity defined sample. Of the 10 X-ray faint, MIR bright
clusters defined by flux, 7 are present in the corresponding
luminosity defined sample.

\begin{figure}
\centering
\psfig{figure=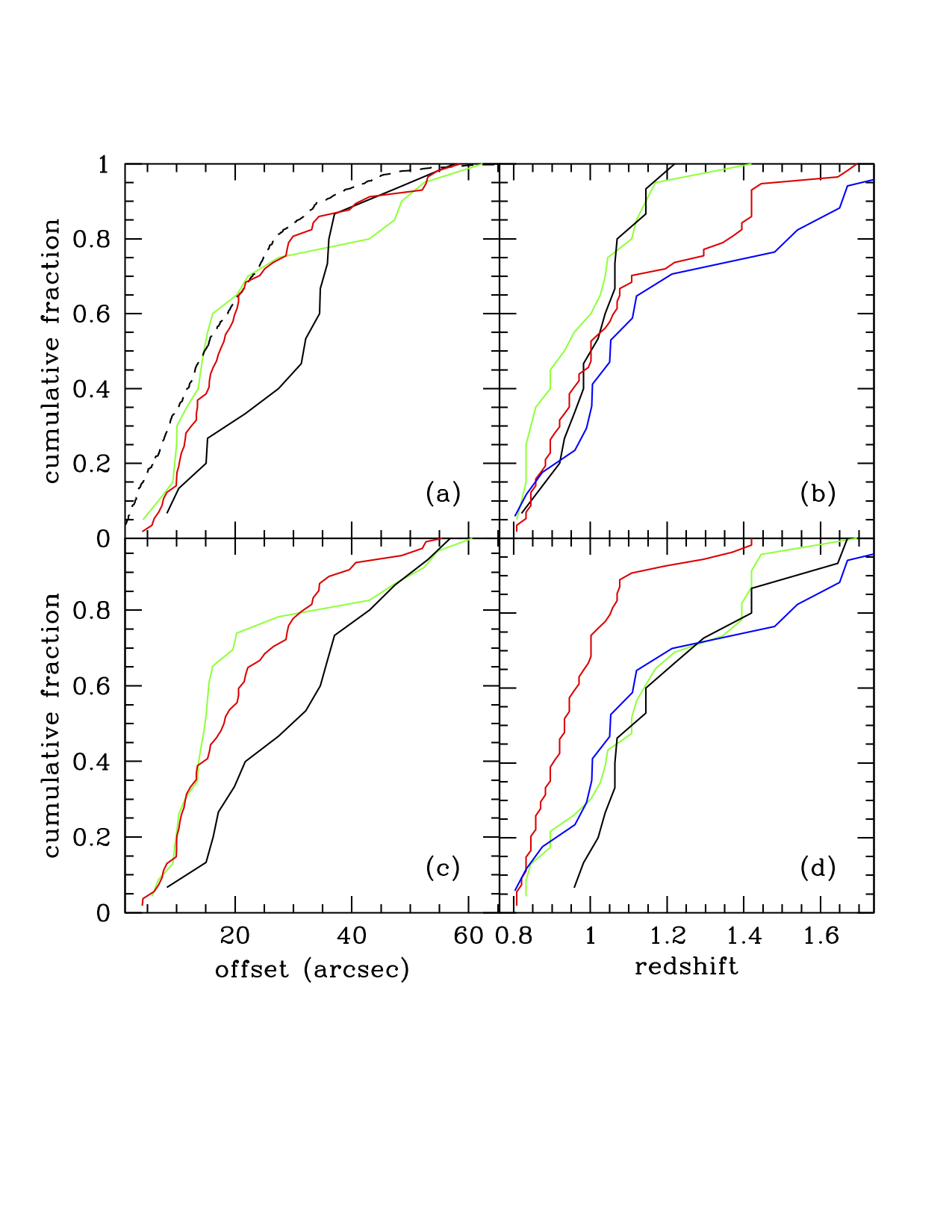,width=3.5in,angle=0.0}
\caption{The cumulative fraction distribution in both offset
  (BCG-barycentre position) and redshift of each cluster sub-sample:
  XMM-LSS (blue), X-ray bright SpARCS (green), X-ray faint MIR faint
  SpARCS (red), X-ray faint, MIR bright (black). Panels: a) offset
  distribution of cluster sub-samples selected on the basis of
  aperture flux measurements, b) redshift distribution of cluster
  sub-samples selected on the aperture flux measurements. Panels c)
  and d) follow panels a) and b) for cluster sub-samples selected on
  the basis of aperture luminosity measurements. The black dashed line
  in panel (a) indicates the BCG-barycentre offset distribution
  expected from a 22\arcsec\ 1D Gaussian random error in the
  barycentre centroid values (see Section \ref{surface_brightness} for
  more details).}
\label{offset_z_dist}
\end{figure}

Figure \ref{offset_z_dist} shows the cumulative fraction distribution
in positional offset (BCG-barycentre position) and redshift of each
cluster sub-sample selected either on the basis of aperture flux or
aperture luminosity measurements.  A number of trends are apparent,
notably that the X-ray faint, MIR faint cluster sub-sample is composed
of distant, i.e. higher redshift, intrinsically luminous clusters when
selected by aperture flux and nearby, i.e. lower redshift,
intrinsically faint clusters when selected by luminosity.  In each case
the sub-sample of X-ray faint, MIR bright clusters, whether selected
by flux or luminosity, displays a distribution of position offsets
between the BCG and barycentre position dominated by higher values
than other cluster sub-samples.
Furthermore, this conclusion holds whether the selection of X-ray
faint, MIR bright clusters is performed using either $r^\prime-3.6$ or
$z^\prime-3.6$ colour and either BCG or barycentre centroids for
aperture measurements.

\subsection{Cumulative MIR angular surface brightness profiles}
\label{surface_brightness}

\begin{figure*}
\centering
\psfig{figure=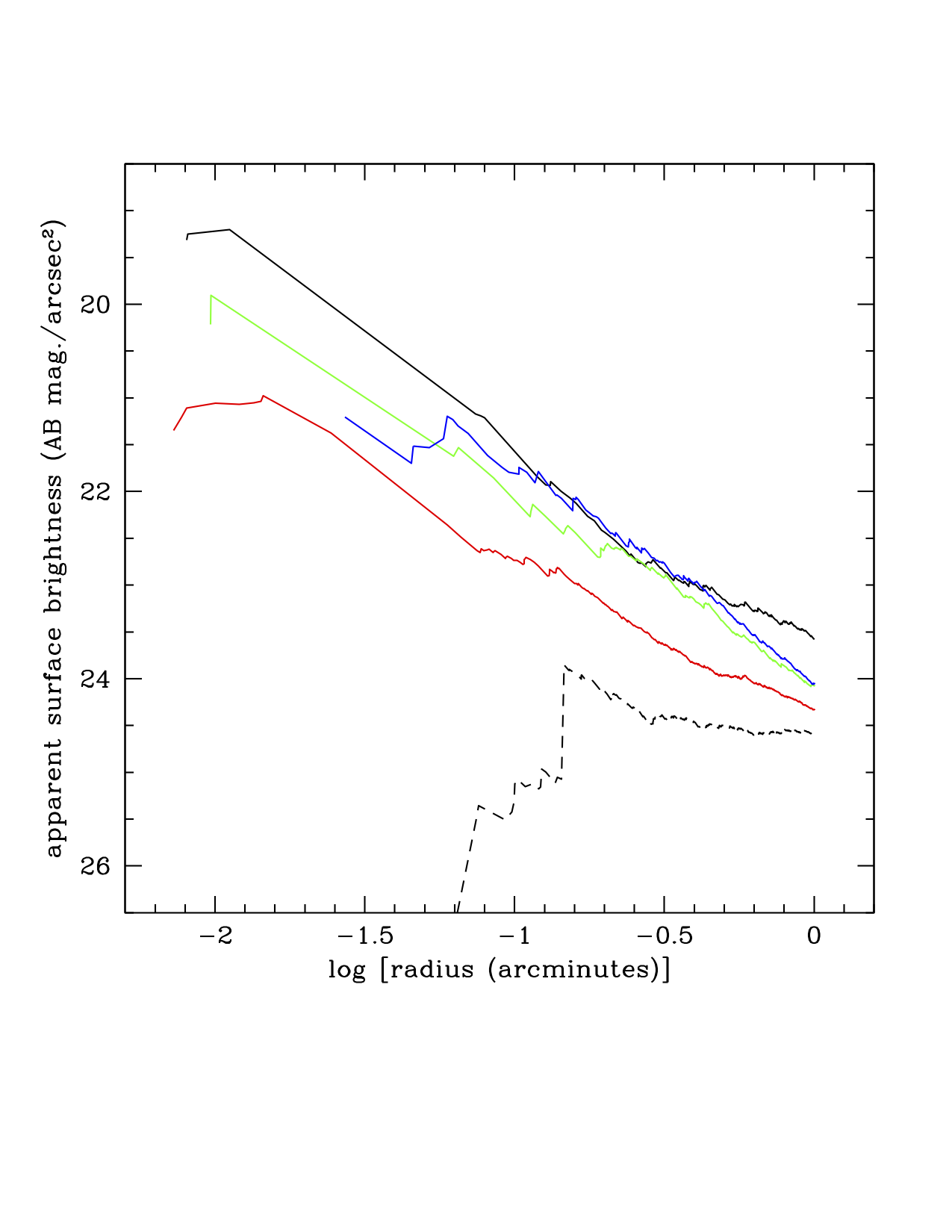,width=3.0in,angle=0.0}
\psfig{figure=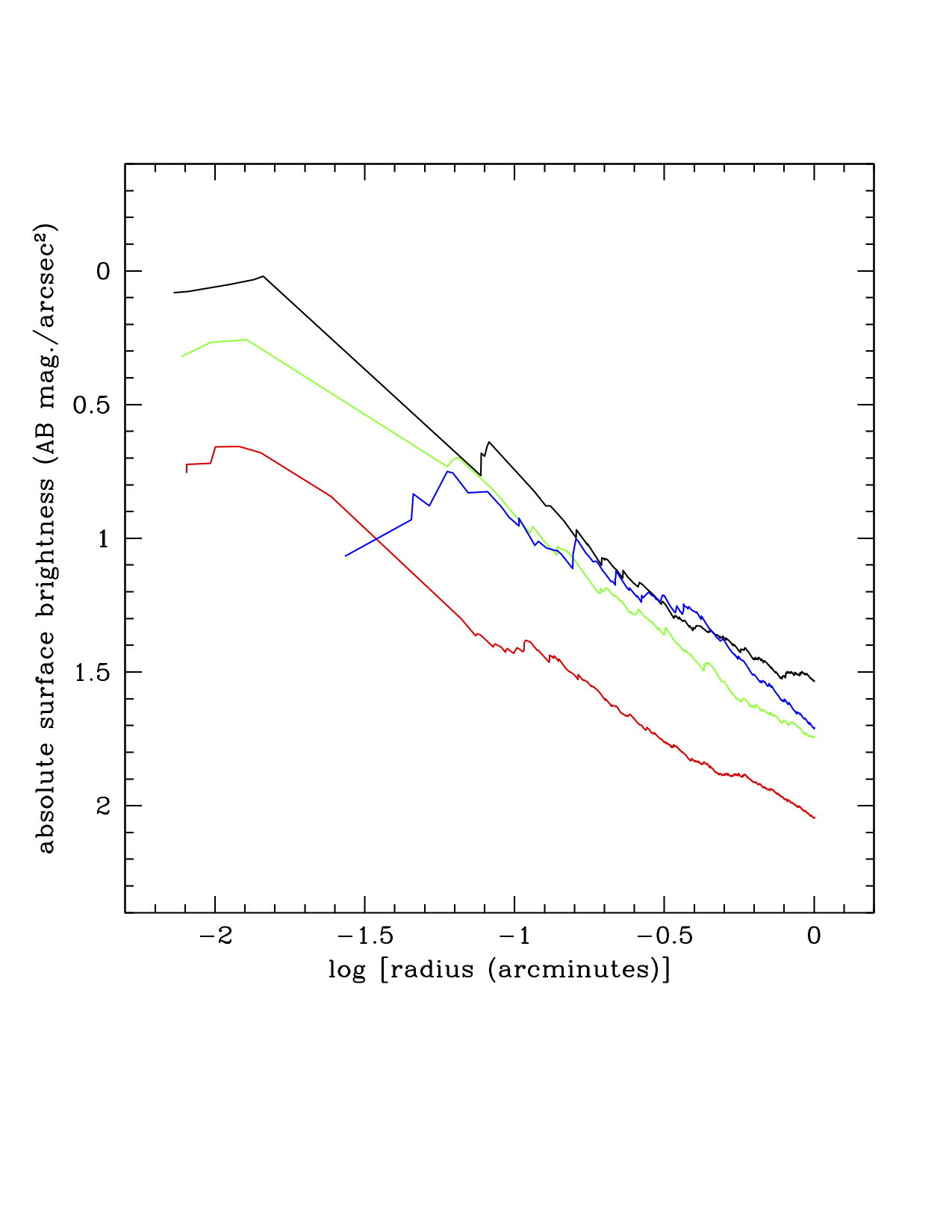,width=3.0in,angle=0.0}
\caption{The cumulative angular MIR surface brightness distribution of
  each cluster sub-sample: XMM-LSS (blue), X-ray bright SpARCS
  (green), X-ray faint MIR faint SpARCS (red), X-ray faint, MIR bright
  (black). Left panel: Surface brightness computed using flux. The
  black dashed line indicates the average cumulative surface
  brightness distribution computed from 100 randomly placed
  apertures. Right panel: Surface brightness computed using
  luminosity. Note that no background is displayed in this panel as it
  is not possible to apply a conversion between flux and luminosity
  given the unknown redshift distribution of the background galaxies.}
\label{cumul_sb}
\end{figure*}

The average MIR angular surface brightness distribution within each
cluster sub-sample was computed using the relation
\begin{equation}
{
\mu (r_i) = \frac{\sum\limits_{i=1}^N f_{3.6\mu m,i}}{4 \pi r_N^2},
}
\label{eqn_sb}
\end{equation}
where $f_{3.6\micron,i}$ and $r_i$ are respectively the
3.6\micron\ flux and angular separation from the appropriate cluster
centroid of a list of $N$ galaxies within each cluster sub-sample
ordered by increasing $r_i$. Once again, the X-ray centroid is used as
the reference position for XMM-LSS clusters and the BCG position is
used for SpARCS. Note that to avoid infinite values of central surface
brightness in the case where the BCG position is used we apply a
softening radius to Equation \ref{eqn_sb} of the form $r_N = r_N+r_S$
with $r_S$ equal to $1\farcs8$.  We also note that the above
cumulative formalism generates qualitatively the same results as a
differential approach which computes the surface brightness in radial
bins about each cluster centre yet avoids an arbitrary choice of
radial bins.  The surface brightness expressed in terms of rest-frame
$K$-band luminosity per unit area can also be computed in a
straightforward manner by replacing the flux of each colour-selected
candidate cluster galaxy along the line-of-sight with the luminosity
computed with the appropriate distance modulus and $k$-correction
discussed in Section \ref{sec_flux_flux}. Figure \ref{cumul_sb}
displays the resulting surface flux and luminosity profiles for each
cluster sub-sample.
Taken together, the figures indicate that both X-ray selected and
either X-ray bright or MIR-bright colour-selected clusters display
similar projected distributions of galaxy light respectively about
either the X-ray or BCG location.  X-ray faint, MIR-faint SpARCS
clusters display similar profiles yet with lower normalisation.  Note
that this procedure does not account for the background of non-cluster
galaxies along the line-of-sight. However, each distribution tends to
an asymptotic background value at large radius.  In the left panel of
Figure \ref{cumul_sb} one notes that all distant cluster projected
surface brightness distributions are clearly different from the
average distribution of background galaxies within a 1 arcminute
radius aperture.

\begin{figure*}
\centering
\psfig{figure=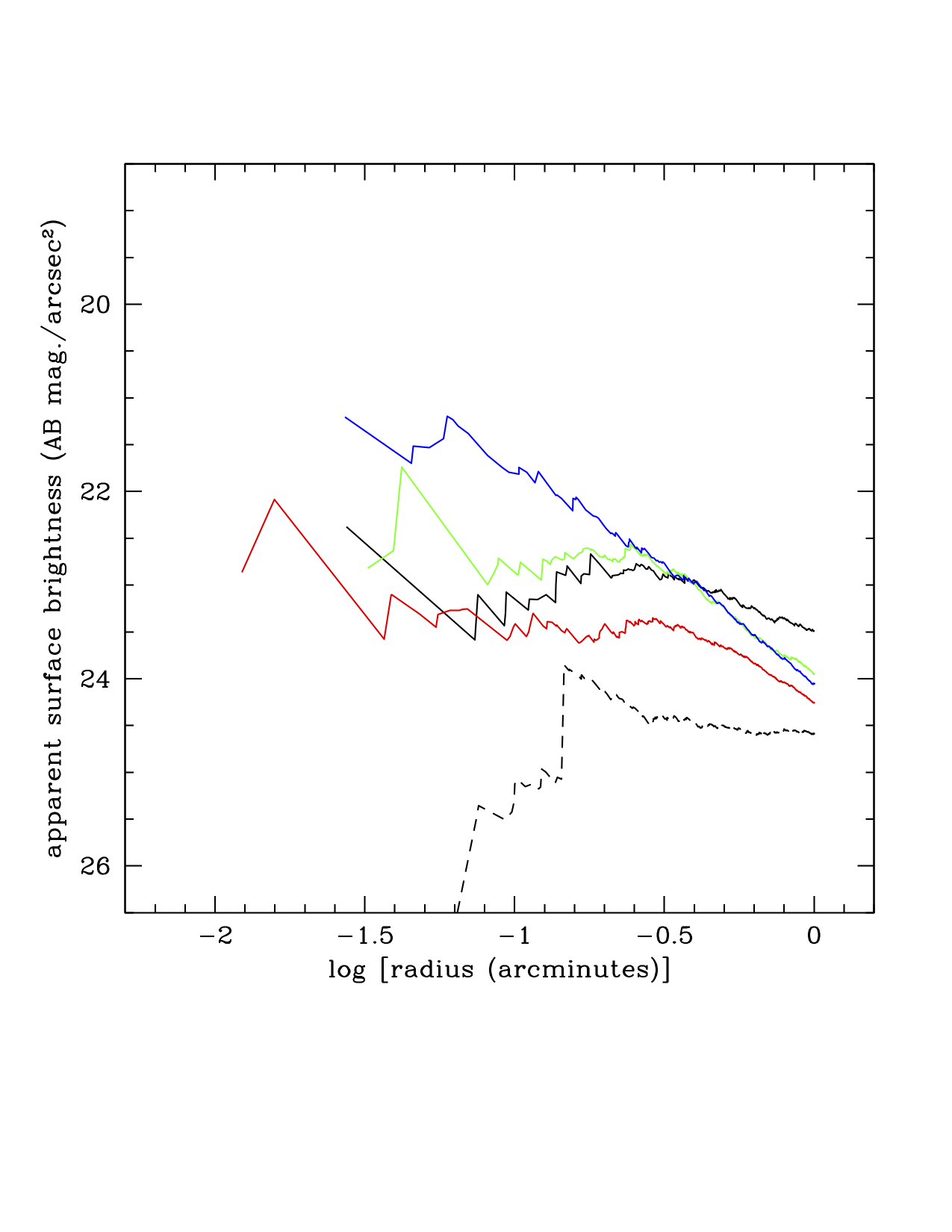,width=3.0in,angle=0.0}
\psfig{figure=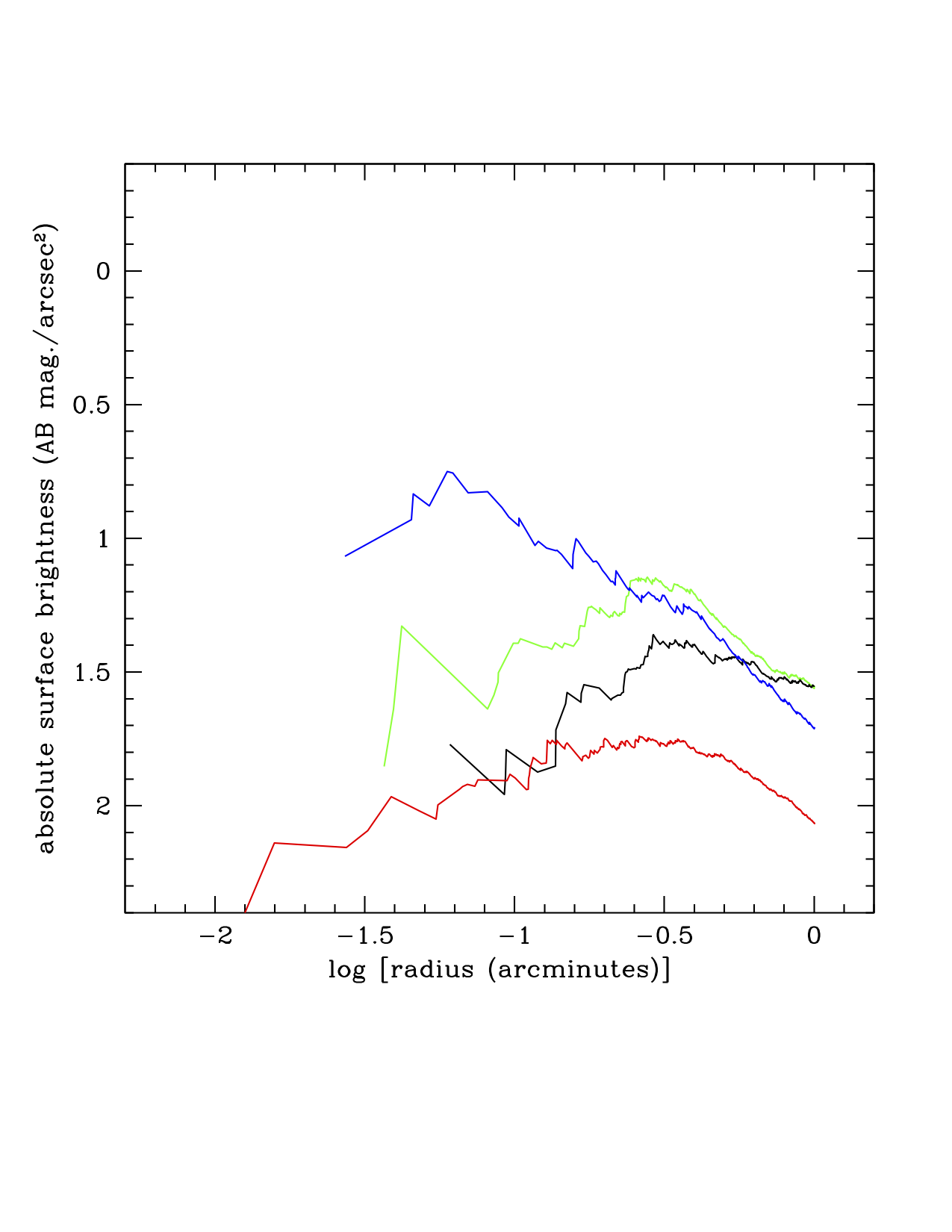,width=3.0in,angle=0.0}
\caption{The cumulative angular MIR surface brightness distribution of
  each cluster sub-sample: XMM-LSS (blue), X-ray bright SpARCS
  (green), X-ray faint MIR faint SpARCS (red), X-ray faint, MIR bright
  (black). Centroids for SpARCS clusters employ the barycentre
  position. Note that the axis scale employed in this figures is the
  same as Figure \ref{cumul_sb}.}
\label{cumul_sb_bary}
\end{figure*}

We repeated the above analysis using the barycentre position of each
SpARCS cluster instead of the BCG location. The results are plotted in
Figure \ref{cumul_sb_bary} and indicate that each sub-sample of SpARCS
clusters displays a central deficit of light compared to the analysis
using the BCG location.  

\begin{figure}
\centering
\psfig{figure=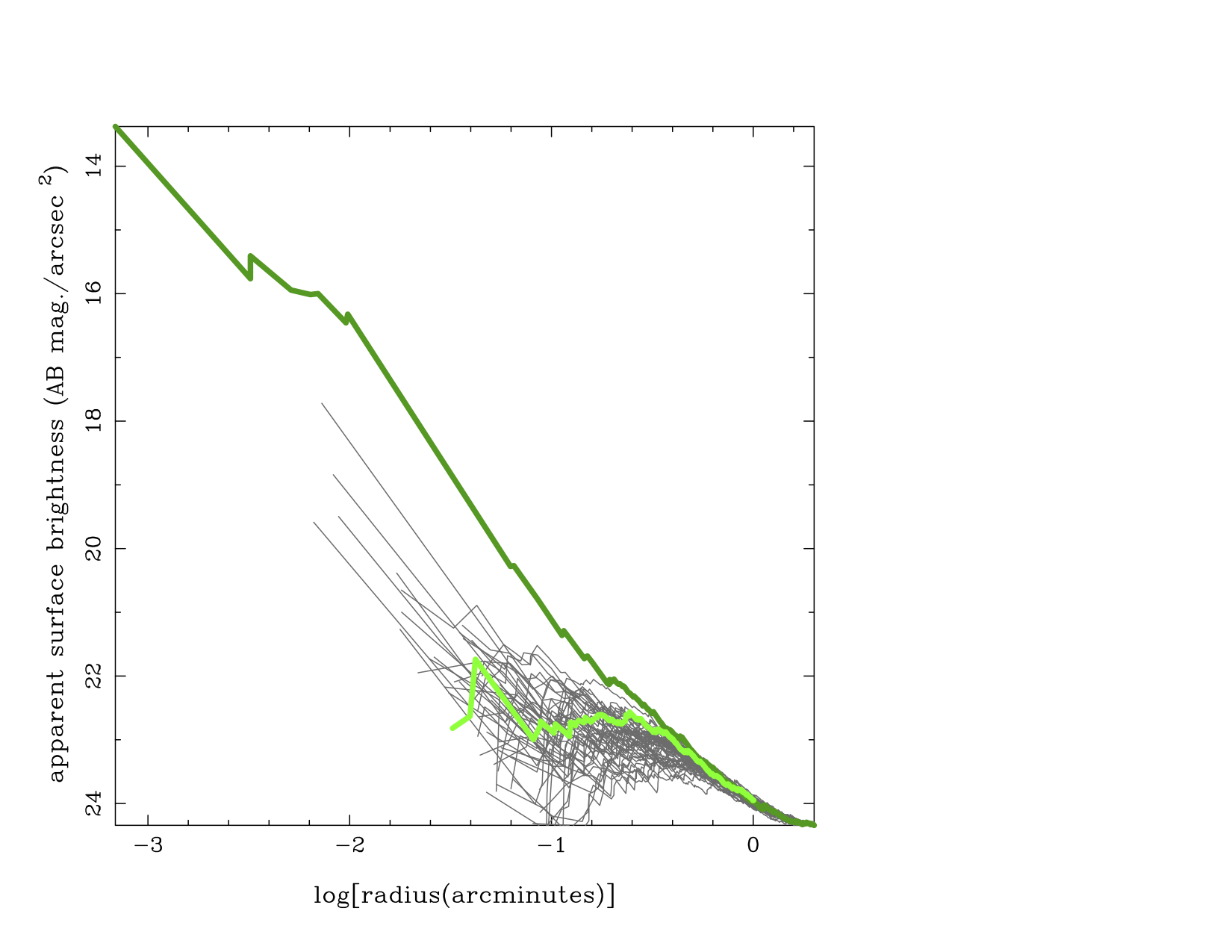,width=4.2in,angle=0.0}
\caption{A detailed comparison of the cumulative surface brightness
  distributions of the SpARCS X-ray bright clusters employing BCG
  (dark green) and barycentre (light green) centroid values. The set
  of grey lines indicate 50 simulations of the stacked surface
  brightness calculation in which a random Gaussian offset of
  $\sigma=15$\arcsec\ is applied to each cluster BCG R.A. and
  dec. centroid prior to stacking.}
\label{cen_sim}
\end{figure}

We investigated whether this apparent deficit was due to large errors
in the barycentre centroid values relative to the BCG positions
measured for the SpARCS clusters.  Such centroid errors, based upon
the average position of candidate cluster members, are a persistent
feature of galaxy overdensity cluster finding algorithms
\citep[e.g.][]{lin2004,rozo2014b,oguri2017}. We restricted our
analysis to the SpARCS X-ray bright clusters assuming that these
represent bona-fide clusters where the BCG location is close to the
true centroid of each cluster.  We applied a random Gaussian offset to
the measured BCG right ascension and declination of each cluster and
recomputed the stacked surface brightness profile as described
above. We repeated this process $N$ times to generate an ensemble of
surface brightness distributions at each specified offset (Figure
\ref{cen_sim}).

The results indicate that the SpARCS X-ray bright barycentre surface
brightness distribution is consistent with that generated employing
BCG positions convolved with a 1D centroid error in R.A. and dec. of
$\sigma=15$\arcsec, corresponding to an error of 22\arcsec\ in radius.
Figure \ref{offset_z_dist} (panel (a)) indicates that this error model
describes a large component, though not all, of the observed
distribution of BCG-barycentre offset values for the SpARCS X-ray
bright (and X-ray faint, MIR faint) sub-samples.

The X-ray faint, MIR bright SpARCS clusters display a distribution of
centroid offsets in excess of this centroid error model and likely
indicates an additional, intrinsic offset distribution.  In studies at
lower redshifts, large positional offsets between the BCG location and
other measures of cluster centroid (e.g. X-ray location or average
member galaxy location) are often taken as an indicator of a cluster
displaying the effects of incomplete virial relaxation
\citep[e.g.][]{sanderson2009,lavoie2016}.

It may be that the location of X-ray faint, MIR bright SpARCS clusters
on the X-ray versus MIR aperture brightness diagrams shown in Section
\ref{sec_flux_flux} is explained by such disturbance/virialisation
arguments.  However, further evidence must be considered before
reaching a conclusion, namely performing a visual assessment of
individual clusters in each sub-sample and considering stacked,
two-dimensional X-ray images of each sub-sample.

\subsection{Visual assessment}

\begin{figure*}
\centering
\psfig{figure=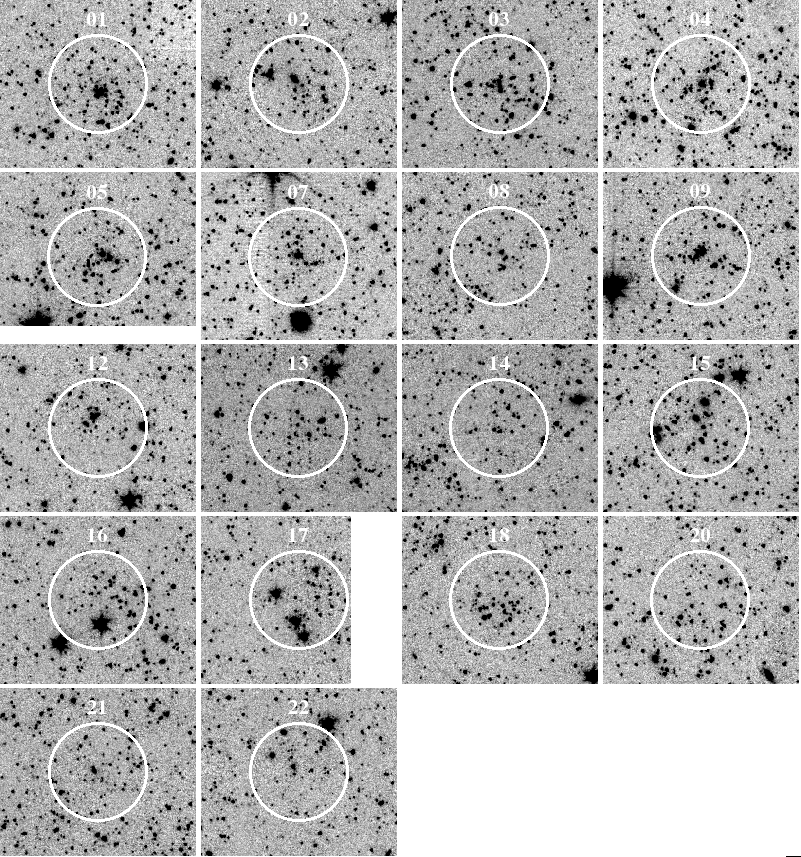,width=7.0in,angle=0.0}
\caption{Spitzer/IRAC 3.6\micron\ images of each cluster in the
  XMM-LSS sample. The white circle in each image displays the
  1\arcmin\ circular aperture applied to each cluster. Each image is
  oriented north up and east left.}
\label{image_willis}
\end{figure*}

\begin{figure*}
\centering
\psfig{figure=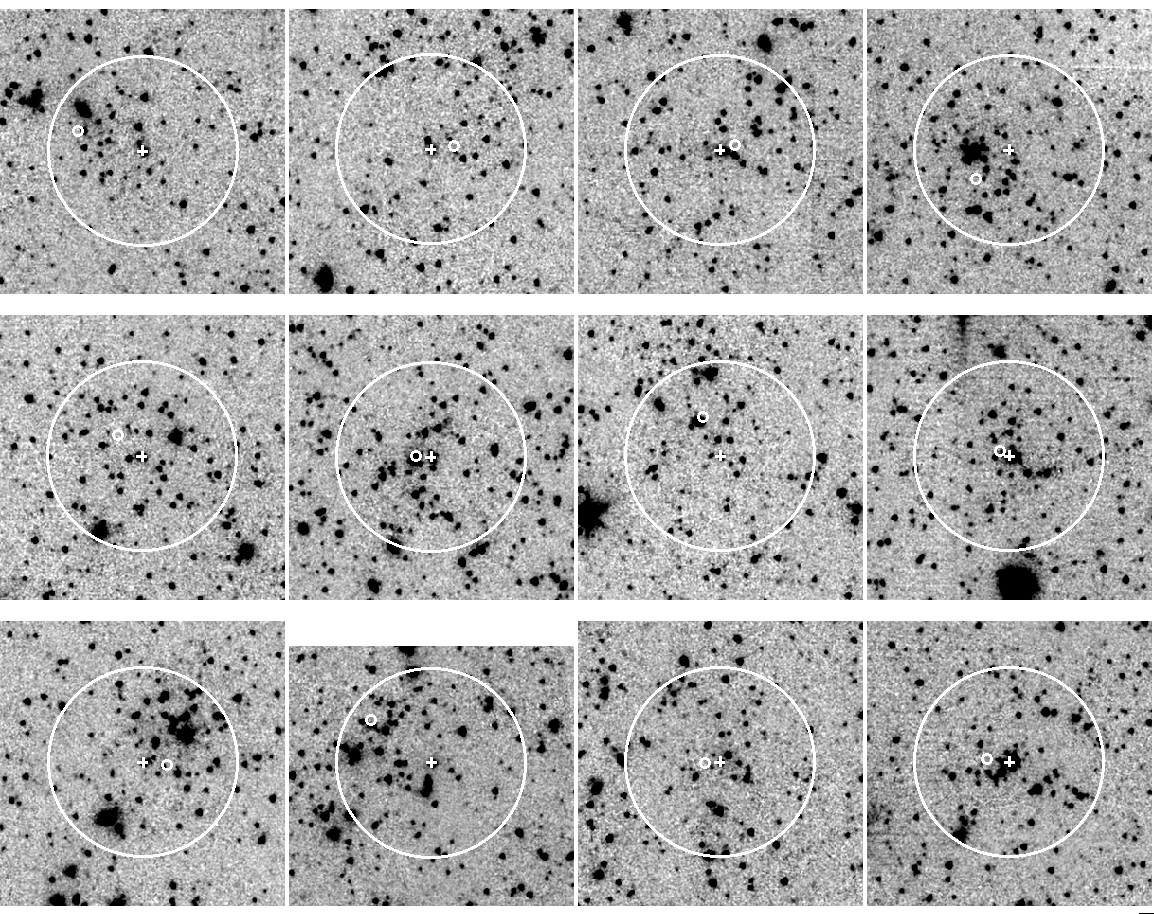,width=7.0in,angle=0.0}
\caption{Spitzer/IRAC 3.6\micron\ images of each cluster in the SpARCS
  X-ray bright sub-sample. The white circle in each image displays the
  1\arcmin\ circular aperture applied to each cluster and is centered
  on the BCG position. The BCG and barycentre positions of each
  cluster are indicated respectively by the cross and the circle. Each
  image is oriented north up and east left.}
\label{image_xbright}
\end{figure*}

\begin{figure*}
\centering
\psfig{figure=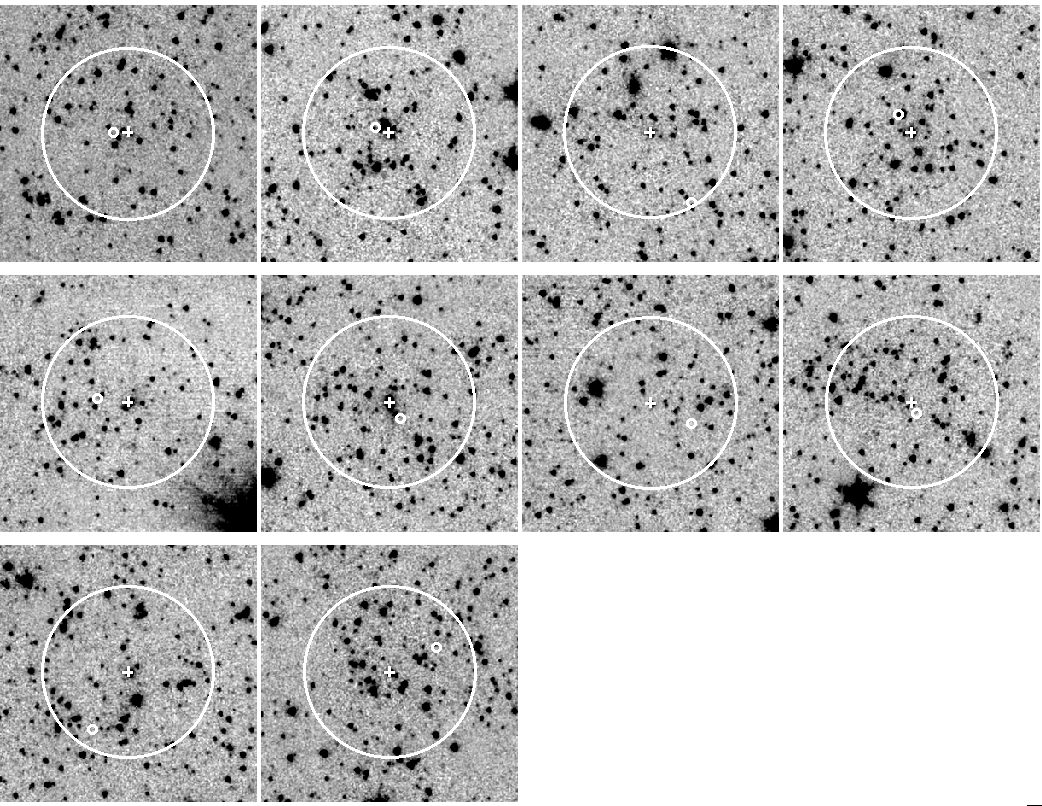,width=7.0in,angle=0.0}
\caption{Spitzer/IRAC 3.6\micron\ images of each cluster in the SpARCS
  X-ray faint, MIR bright sub-sample. The white circle in each image
  displays the 1\arcmin\ circular aperture applied to each cluster and
  is centered on the BCG position. The BCG and barycentre positions of
  each cluster are indicated respectively by the cross and the
  circle. Each image is oriented north up and east left.}
\label{image_xfaint_mirbright}
\end{figure*}

\begin{figure*}
\centering
\psfig{figure=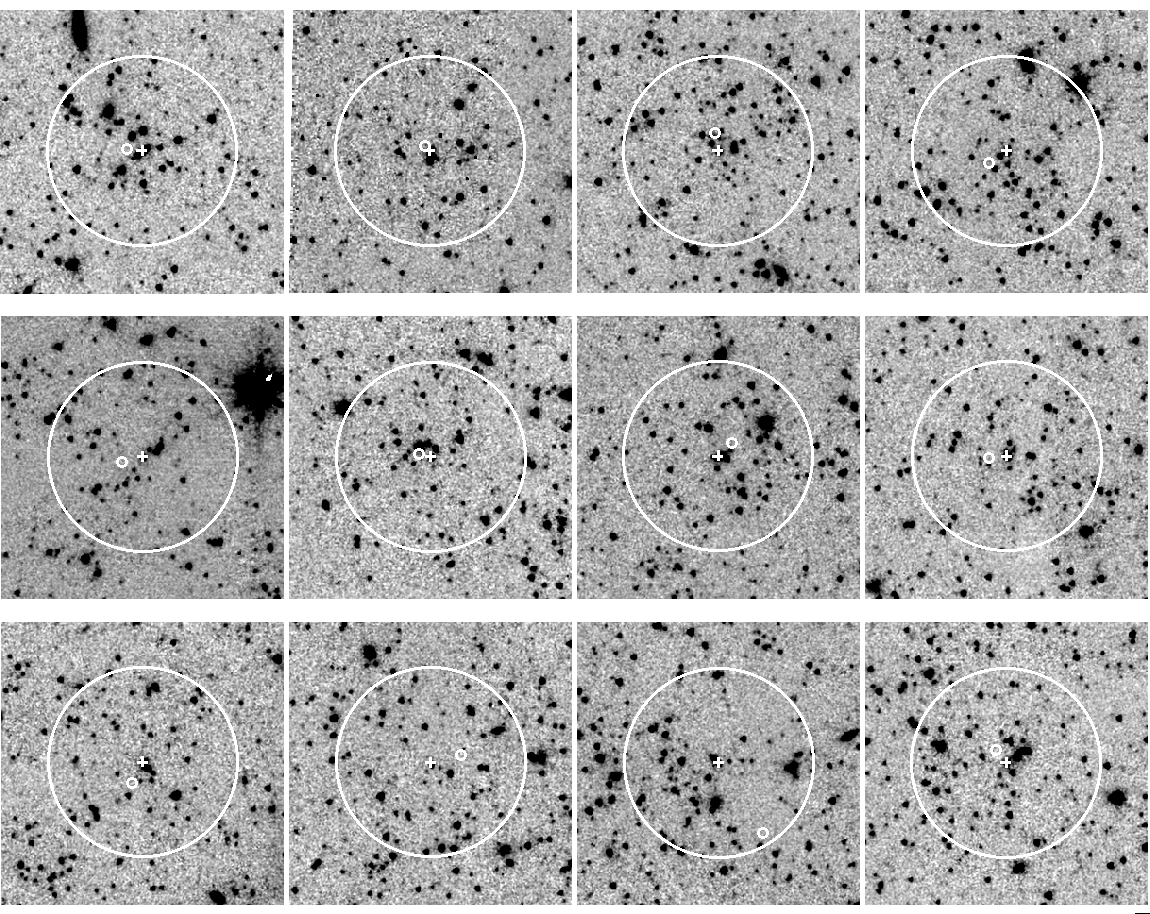,width=7.0in,angle=0.0}
\caption{Spitzer/IRAC 3.6\micron\ images of a subset of clusters from
  the SpARCS X-ray faint, MIR faint sub-sample.  The top, middle and
  bottom rows show typical clusters at low, middle and high redshift
  within this sub-sample.  The white circle in each image displays the
  1\arcmin\ circular aperture applied to each cluster and is centered
  on the BCG position. The BCG and barycentre positions of each
  cluster are indicated respectively by the cross and the circle. Each
  image is oriented north up and east left.}
\label{image_xfaint_mirfaint}
\end{figure*}

It is sensible to determine whether the trends in radial surface
brightness presented in Section \ref{surface_brightness} are supported
by a visual assessment of individual clusters in each
sub-sample. Figures \ref{image_willis}, \ref{image_xbright},
\ref{image_xfaint_mirbright} and \ref{image_xfaint_mirfaint} display
Spitzer/IRAC 3.6\micron\ images of clusters drawn from the XMM-LSS,
SpARCS X-ray bright, SpARCS X-ray faint, MIR bright and SpARCS X-ray
faint, MIR faint samples respectively.

Although a visual assessment can only deliver qualitative information,
it is clear that the images of clusters in each sub-sample support the
surface brightness trends presented in Section
\ref{surface_brightness} with both the XMM-LSS and SpARCS clusters
appearing as centrally concentrated systems of galaxies.

The sub-sample of SpARCS X-ray faint, MIR faint clusters presents a
range of appearances, largely consistent with their low MIR aperture
fluxes.  There is some evidence for central concentrations of galaxies
in the 3.6\micron\ images.  However, there are also numerous examples,
particularly at higher redshift, of images sparsely populated by
galaxies where no conclusive statement can be made on the basis of
visual inspection.

\subsection{Stacked colour magnitude diagrams}

The creation of stacked colour magnitude diagrams for each cluster
sub-sample provides an opportunity to assess whether each represents a
population of galaxies drawn from a narrow range of star formation
histories.  In particular the presence, location and width of the
characteristic cluster red sequence provides measure of the average
evolved galaxy population within each cluster sub-sample.

Photometry corresponding to 3.6\micron\ magnitudes and
$r^\prime-3.6\micron$ colours for galaxies located within 1\arcmin\ of
each cluster centroid were obtained.  Photometry for different
clusters in each sub-sample were each transformed from the catalogue
redshift of each cluster to a common reference of $z=1$ by applying an
appropriate $k-$ and distance modulus correction as discussed in
Section \ref{sec_flux_flux}.

\begin{figure}
\centering
\psfig{figure=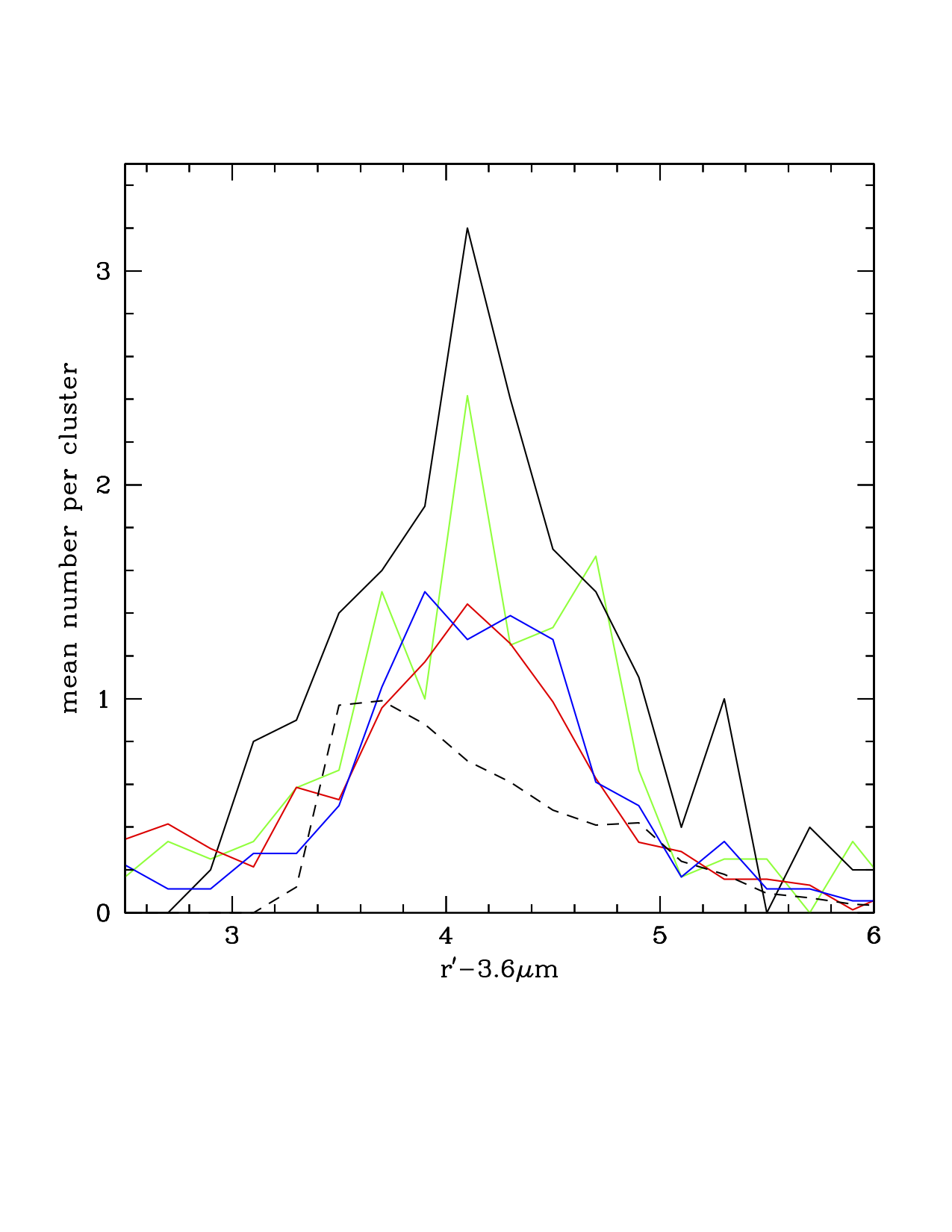,width=3.5in,angle=0.0}
\caption{Average $r^\prime-3.6\micron$ colour histogram for clusters
  in each sub-sample: XMM-LSS (blue), X-ray bright SpARCS (green),
  X-ray faint, MIR faint SpARCS (red), X-ray faint, MIR bright
  (black). The dashed black line shows the scaled colour histogram
  generated by placing 100 1\arcmin\ apertures at random within the
  survey area and extracting all sources satisfying the distant galaxy
  colour cut. Only sources satisfying $[3.6\micron] < 21$ when
  corrected to $z=1$ are displayed. No $k$- or distance modulus
  corrections are applied to the randomly collected source
  photometry. }
\label{cmd_stack_all}
\end{figure}

Figure \ref{cmd_stack_all} isolates the red sequence distribution by
collapsing the colour-magnitude plane along the 3.6\micron\ magnitude
axis to create a colour histogram for each cluster sub-sample.  All
sources displaying a redshift-corrected magnitude $[3.6\micron]<21$ AB
are co-added following this method.  Despite the changing
normalisation of each sub-sample as one proceeds from XMM-LSS clusters
to SpARCS X-ray bright and finally to SpARCS X-ray faint, MIR faint
clusters it is clear that each displays a similar red sequence
distribution, i.e. each contains a red galaxy population of
approximately similar absolute age and scatter.

The comparison indicates that the typical number of red sequence
galaxies located in XMM-LSS clusters is comparable to that found in
the X-ray faint, MIR faint sub-sample of SpARCS clusters, echoing
earlier comparisons between distant X-ray and IR-selected clusters
\citep{foltz2015}.  SpARCS clusters labelled as X-ray bright and X-ray
faint, MIR bright each display more populous red sequence
distributions, marginally in the case of X-ray bright systems more
significantly for X-ray faint, MIR bright clusters (though omitted for
clarity, the typical Poisson error for each data point in Figure
\ref{cmd_stack_all} is approximately 0.25-0.3).

\subsection{Stacked X-ray images}
\label{xstack}

The creation of stacked X-ray images for each cluster sub-sample
permits the average X-ray emission properties of each to be
discussed. Furthermore, the low noise properties of stacked images in
particular permits a sensitive test of the average emission from the
SpARCS X-ray faint sub-samples to be investigated.

\begin{figure*}
\centering
\psfig{figure=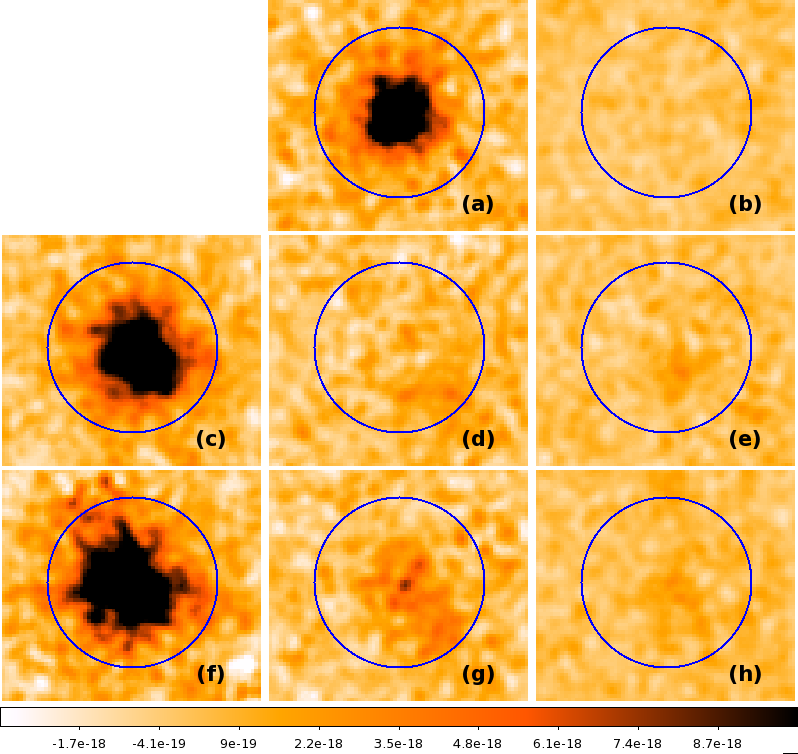,width=7in,angle=0.0}
\caption{Stacked X-ray flux images for each cluster sub-sample. Each
  image is smoothed with a Gaussian kernel of sigma equal to three
  pixels. The scale bar indicates the flux per pixel in units of ergs
  s$^{-1}$ cm$^{-2}$. Panels: a) XMM-LSS clusters, b) Stack of 100
  random positions, c,d,e) SpARCS clusters stacked on the catalogue
  barycentre position, c) X-ray bright, d) X-ray faint, MIR bright, e)
  X-ray faint, MIR faint, f,g,h) SpARCS clusters stacked on the
  catalogue BCG position, f) X-ray bright, g) X-ray faint, MIR bright,
  h) X-ray faint, MIR faint. The blue circle in each panel represents
  the 1\arcmin\ radius aperture used to measure individual cluster
  X-ray fluxes.}
\label{xray_stack_flux}
\end{figure*}

\begin{figure*}
\centering
\psfig{figure=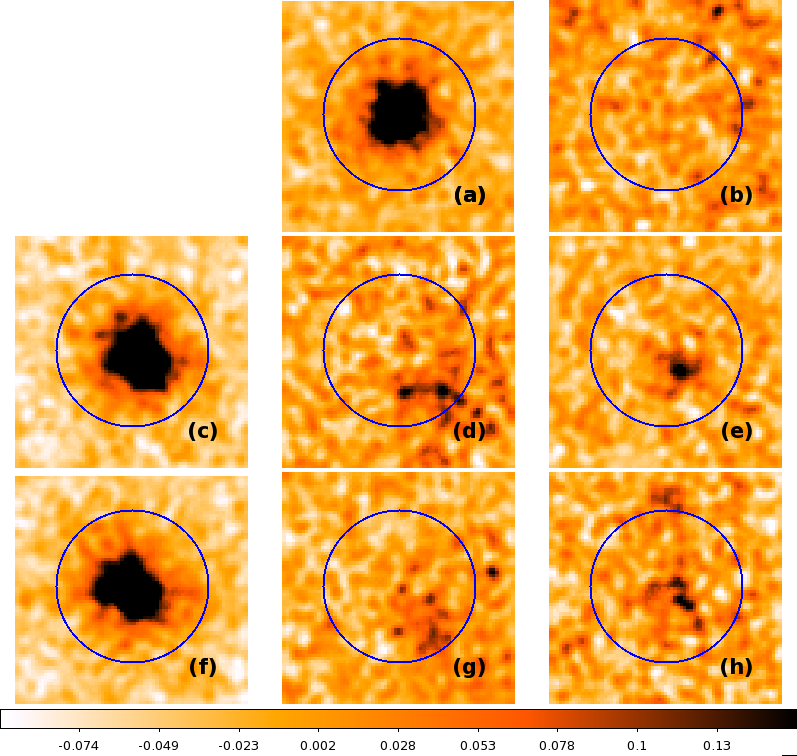,width=7in,angle=0.0}
\caption{Stacked X-ray luminosity images for each cluster
  sub-sample. The caption information is the same as for Figure
  \ref{xray_stack_flux} with the exception that the scale bar
  indicates the luminosity per pixel in units of $10^{41}$ ergs
  s$^{-1}$.}
\label{xray_stack_lum}
\end{figure*}

We examine stacked images of the cluster sub-samples using the data
from the XMM-LSS survey. In brief, for each sub-sample and for each
EPIC detector we follow this procedure:
\begin{enumerate}
 \item Extract a 2\arcmin\ radius EPIC image and corresponding
   exposure map for each cluster in the [0.5-2]~keV band.
 \item Create a background map for each field by fitting a two
   component model to source masked EPIC images. In this way, the
   effects of spatial variation in the background are taken into
   account.
 \item Mask out all point sources in each image, exposure and
   background maps. Point source locations are obtained from the
   XMM-LSS pipeline.
 \item Sum each of the EPIC images to produce a stacked image. Also
   sum each of the individual exposure and background maps. In this
   step, the MOS exposure maps are weighted according to the MOS/PN
   response ratio. The relative sensitivity of the MOS and PN
   detectors is calculated with {\tt XSPEC} using standard on-axis PN
   and MOS response matrices.
\end{enumerate}
The final count-rate image is obtained by subtracting the stacked
background map from the stacked photon image and dividing by the
stacked exposure map. Figures \ref{xray_stack_flux} and
\ref{xray_stack_lum} shows the final images for each sub-sample
stacked by either flux or luminosity. As a test of our flux stacking
procedure we also analyzed a set of 100 randomly selected positions
from the XMM-LSS region in an identical way as the real cluster
positions. Luminosity stacks are created from flux stacks employing an
energy conversion factor based upon the mean redshift of each
sub-sample.

Highly significant X-ray emission is seen in both the XMM-LSS and
X-ray bright SpARCS sub-samples when compared to the image created by
combining 100 random locations.  Weaker emission is also detected in
the stacked image corresponding to the X-ray faint, MIR bright
sub-sample.  It is apparent that the spatial distribution of X-ray
emission in the SpARCS sub-samples is associated more closely with the
cluster BCG position compared to the barycentre position.  This is
indicated visually in the stacks for the X-ray bright and X-ray faint,
MIR bright sub-sample where more compact, centrally-peaked X-ray
emission is generated when stacking on the BCG position.  The stacked
image for the X-ray faint, MIR faint SpARCS sub-sample displays only
marginal X-ray emission in excess of random when stacking on the BCG
position \---\ a result consistent with the low overall aperture X-ray
flux measurements for this sub-sample.  Stacking in luminosity largely
confirms these trends yet with additional substructure present in the
stacked images. This is attributed to individual distant sources in
each sub-sample for which a large correction is obtained when
converting pixel values from count rate to luminosity.

\begin{figure*}
\centering
\psfig{figure=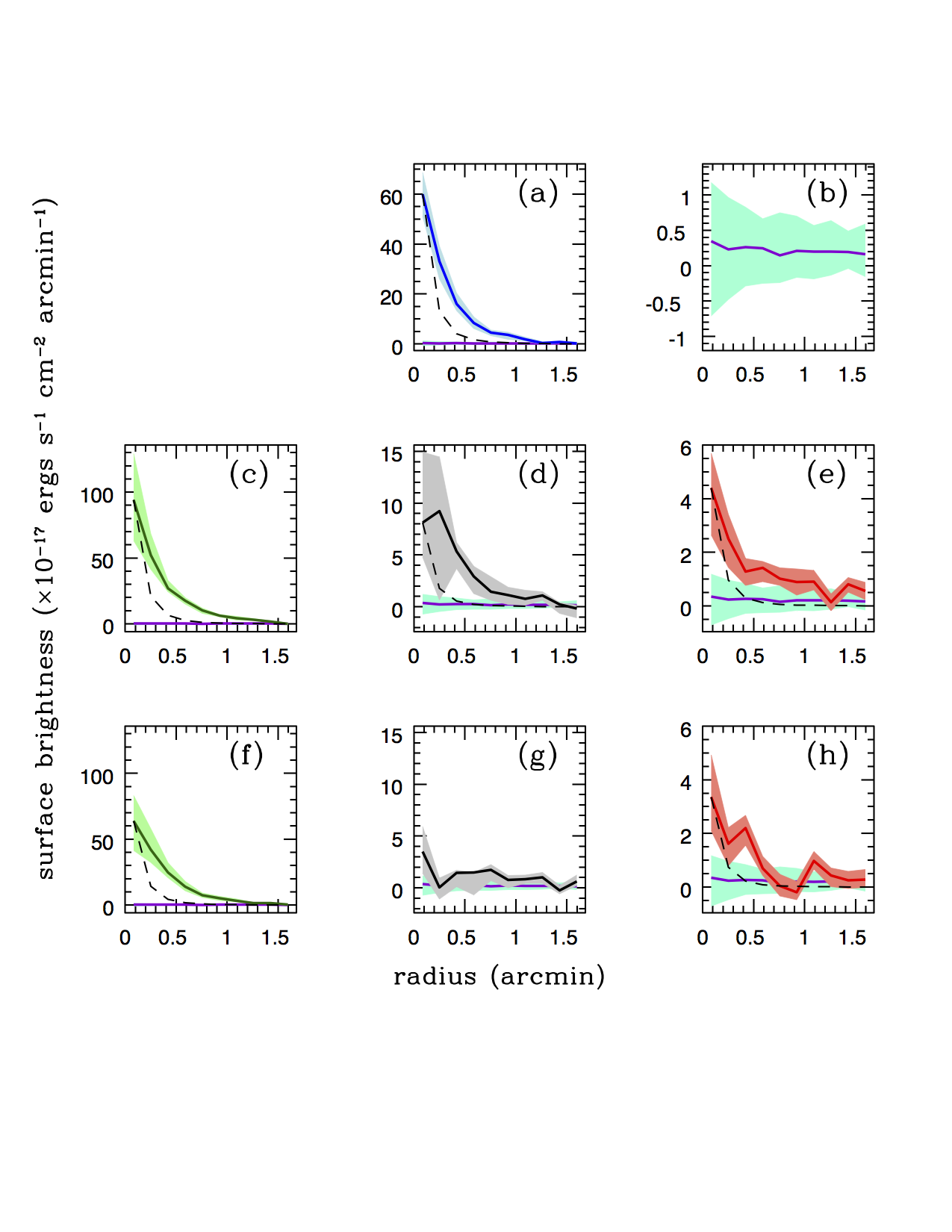,width=6.0in,angle=0.0}
\caption{The differential angular X-ray surface brightness
  distribution in flux of each stacked cluster sub-sample. The
  ordering of the panels follows that of Figure
  \ref{xray_stack_flux}. The shaded area about each solid line
  represents the 68\% confidence interval obtained from 1000 bootstrap
  realisations of each stack. Where shown, the black dashed line
  indicated the XMM PSF scaled to the central surface brightness.}
\label{xsb_flux_all}
\end{figure*}

\begin{figure*}
\centering
\psfig{figure=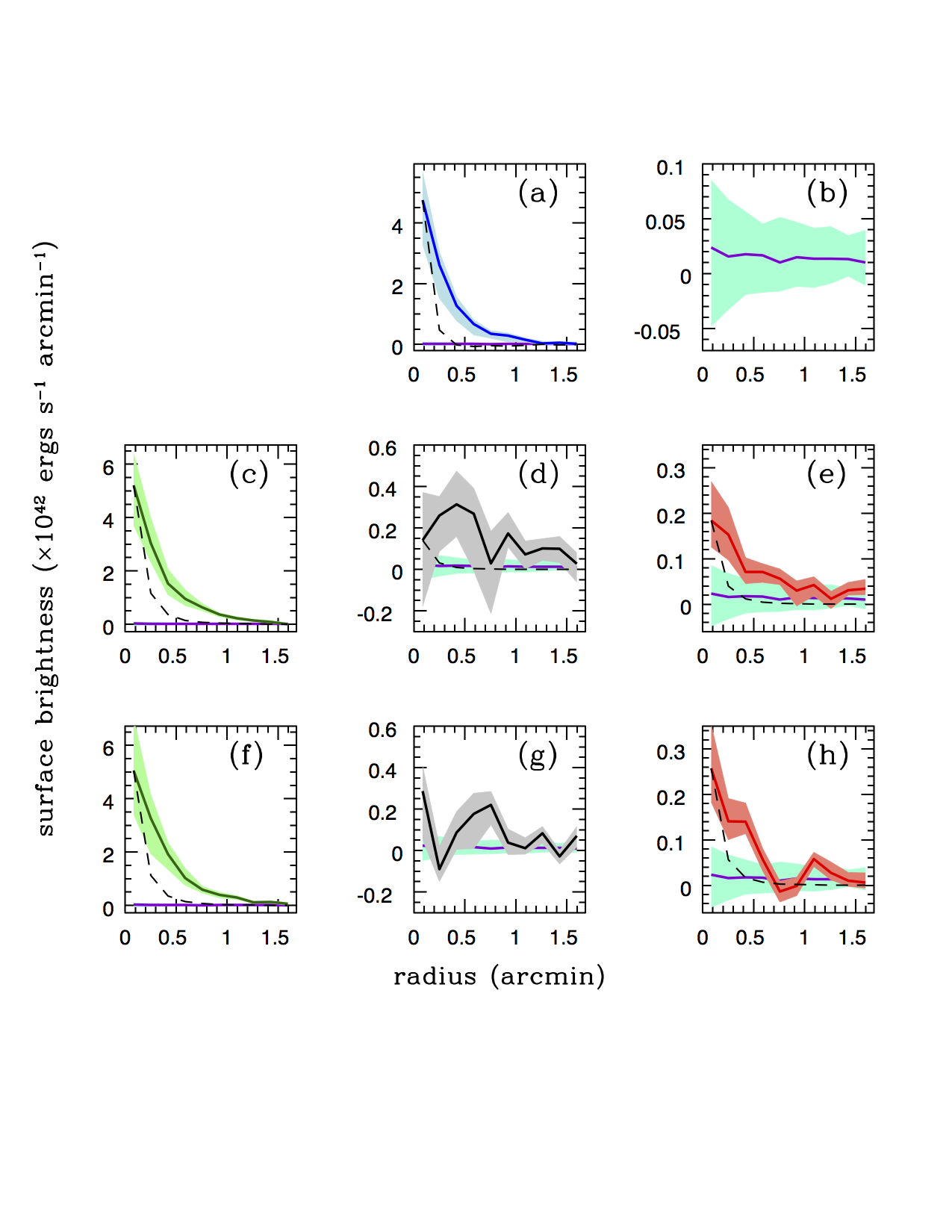,width=6.0in,angle=0.0}
\caption{The differential angular X-ray surface brightness
  distribution in luminosity of each stacked cluster sub-sample. The
  caption information is the same as for Figure \ref{xsb_flux_all}.}
\label{xsb_lum_all}
\end{figure*}

The visual trends noted in the stacked images are reinforced by
inspection of the angular X-ray surface brightness distributions in
each cluster sub-sample shown in Figures \ref{xsb_flux_all} and
\ref{xsb_lum_all}.  The results confirm that the X-ray emission in
each SpARCS sub-sample is consistent with being centred on the BCG and
that the X-ray faint, MIR bright sub-sample displays weak yet
significant extended X-ray emission.  The surface brightness
distributions computed using barycentre centroids are suppressed
relative to those computed using the BCG centroids in a manner
consistent with a barycentre centroid error as discussed in Section
\ref{surface_brightness}.

\begin{figure}
\psfig{figure=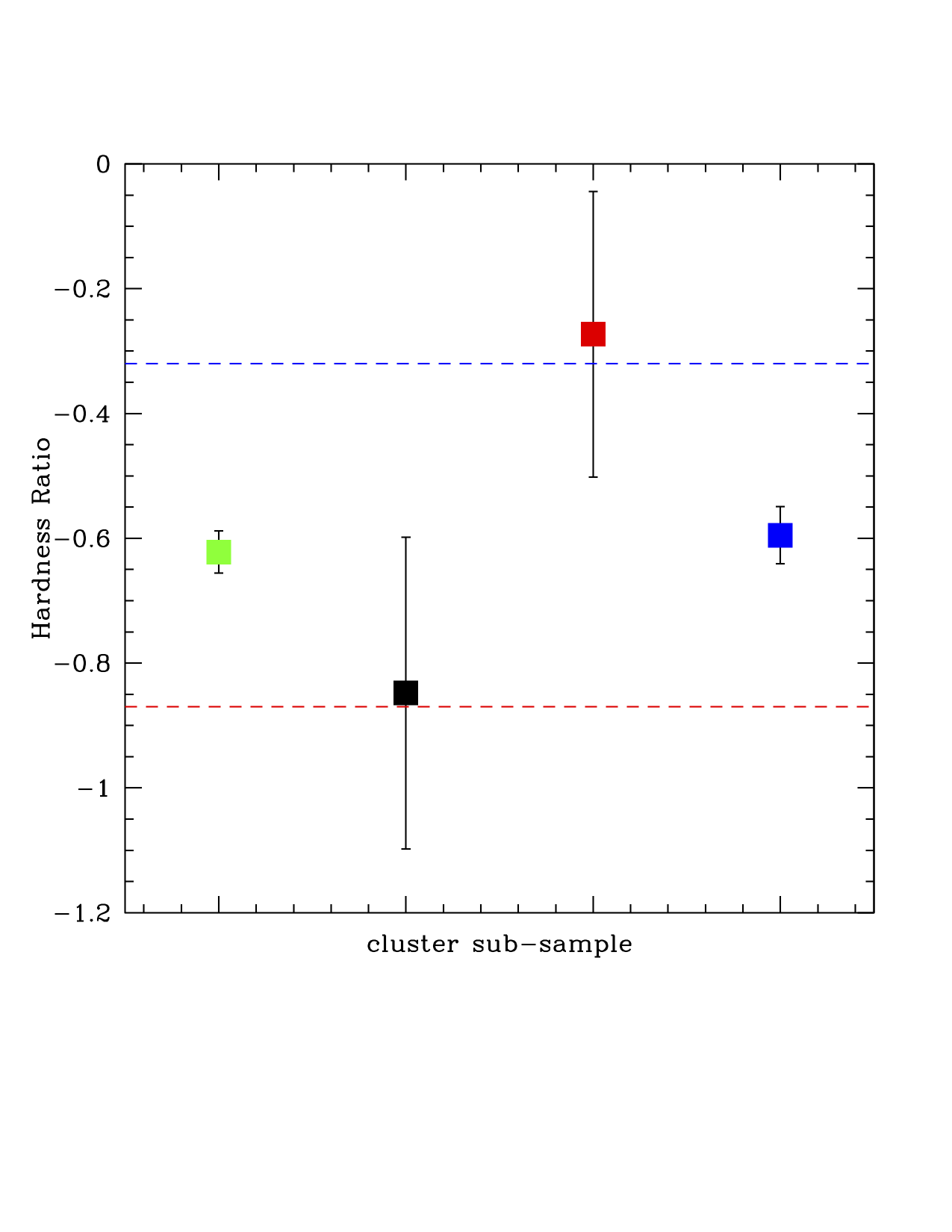,width=3.5in,angle=0.0}
\caption{Hardness ratios computed for each cluster sub-sample: XMM-LSS
  (blue), X-ray bright SpARCS (green), X-ray faint, MIR faint SpARCS
  (red), X-ray faint, MIR bright SpARCS (black). One sigma errors on
  the HR are calculated using a Monte Carlo simulation employing
  posterior probability distribution functions of count-rate within
  the 15\arcsec\ aperture.  The red dashed line indicates the HR
  expected from thermal emission, in this case an absorbed APEC model
  with $T=2$ considered at $z=1$. The blue dashed line indicates the
  HR expected from a simple model of AGN emission, i.e. an absorbed
  power law with index of -2, again considered at $z=1$.}
\label{xhr}
\end{figure}

When interpreting the stacked X-ray emission from each sub-sample,
care must be taken to ensure that the resulting signal is dominated by
extended ICM emission instead of the emission from weak AGN (strong
AGN, i.e. those identified in individual exposures, having been masked
prior to stacking). The stacked surface brightness distribution of
each SpARCS sub-sample is compared to a scaled point spread function
(PSF) appropriate to the combined XMM detectors.  Focussing on the BCG
centroid stacks, panels {\it c/d/e} of Figures \ref{xsb_flux_all} and
\ref{xsb_lum_all} indicate that extended X-ray emission is present in
all three sub-samples.  Following \cite{anderson2013}, we compute the
X-ray hardness ratio as
\begin{equation}
{
HR = \frac{H-S}{H+S},
}
\end{equation}
within the central 15\arcsec\ of each stack using the [0.5-2] keV and
[2-10] keV intervals as the soft and hard band respectively. The
results are plotted in Figure \ref{xhr} and are compared to two,
simple spectral models which respectively represent redshifted thermal
ICM and non-thermal AGN emission (see caption for more details).

Given the simplicity of the model comparison it is perhaps appropriate
only to comment that the XMM-LSS clusters, in addition to the SpARCS
X-ray bright and X-ray faint, MIR bright sub-samples show little
evidence for AGN contamination on the basis of hardness ratio.  The
hardness ratio for the stacked SpARCS X-ray faint, MIR faint
sub-sample is nominally consistent with the simple AGN emission model
presented here, albeit with large errors.  However, the emission
morphology in the stacked SpARCS X-ray faint, MIR faint sample is
clearly extended, leaving the question of the fraction of BCGs in
these clusters that host weak AGN relatively unconstrained.
\cite{webb2015} report that 7/125 or 6\%\ of SpARCS BCGs hosting
bright 24\micron\ sources \---\ the majority of which occurring at $z
\ga 1$ \---\ display IR colours consistent with being dominated by an
AGN.  \cite{martini2013} report that the fraction of X-ray bright AGN
hosted by cluster galaxies in 13 MIR-selected clusters at $1<z<1.5$ is
3\%.  It appears that, although weak emission from AGN associated with
the BCG in each cluster may contribute to each stacked cluster X-ray
image, the level of contamination, at a few percent, is unlikely to be
large. In this sense we consider that the APEC thermal emission model
employed to determine the $k$-correction used to convert between X-ray
flux and luminosity (Section \ref{sec_flux_flux}) remains reasonable.

\begin{figure}
\psfig{figure=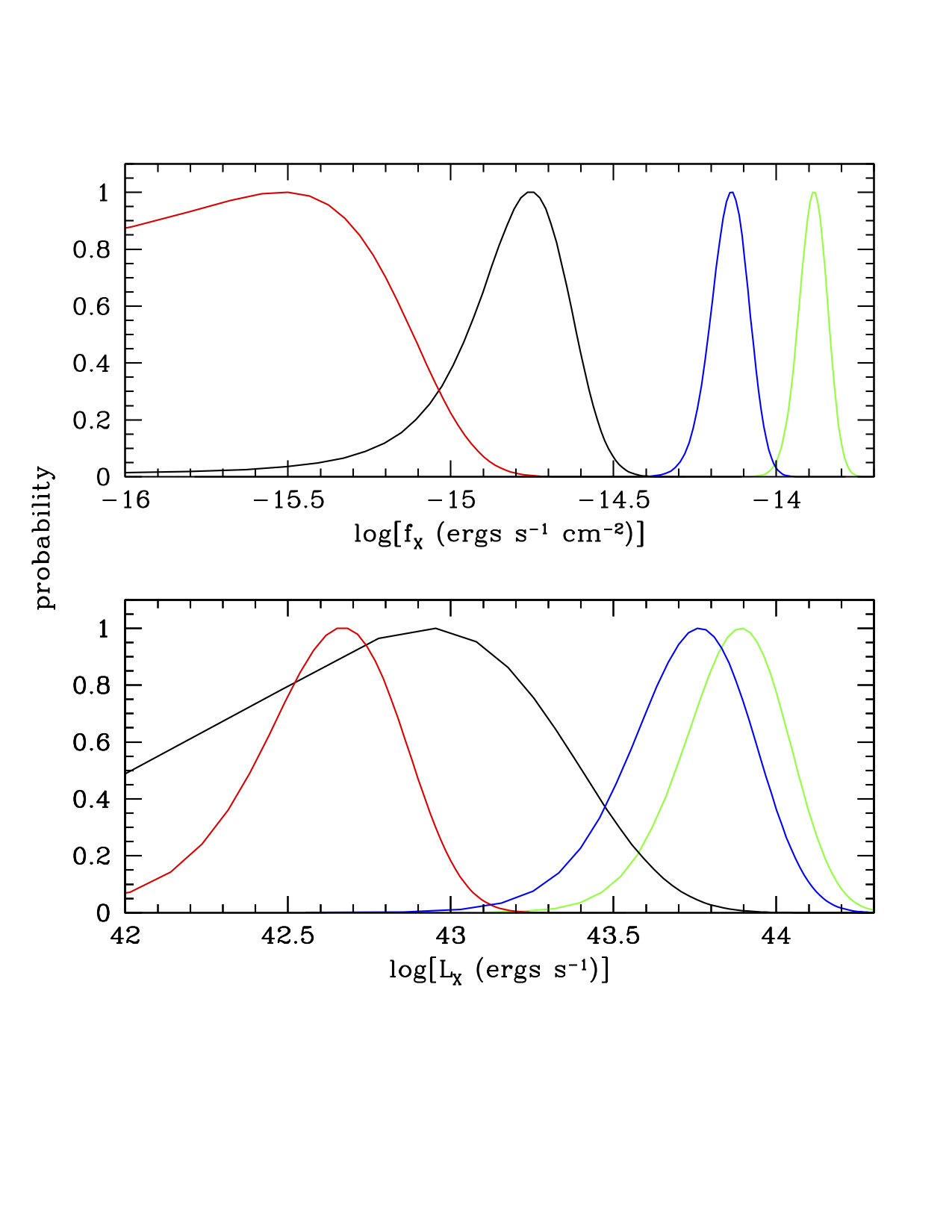,width=3.5in,angle=0.0}
\caption{Posterior probability distributions for aperture flux and
  luminosity measured from the stacked X-ray images: XMM-LSS (blue),
  X-ray bright SpARCS (green), X-ray faint, MIR faint SpARCS (red),
  X-ray faint, MIR bright SpARCS (black).}
\label{stack_ppd}
\end{figure}

Aperture flux and luminosity measurements were computed for the
stacked X-ray images and comparable MIR aperture measurements were
were computed for each cluster sub-sample. Figure \ref{stack_ppd}
displays the posterior probability distributions (PPDs) of each of the
aperture fluxes and luminosities measured from each stacked
image. Stacked aperture measures are plotted in Figure \ref{fig_xflux}
and indicate significant detections for each sub-sample. Values
corresponding to the stacked Xray flux/luminosity represent the mode
of the PPD and the error bars indicate the confidence interval
enclosing 68\% of the distribution. In particular, a significant
(though faint) X-ray detection is obtained for the X-ray faint, MIR
bright stack.

\section{Discussion and Conclusions}
\label{conc}

The comparison of the properties of the XMM-LSS and SpARCS distant
cluster samples has revealed that X-ray bright clusters, whether
detected using X-ray or optical-MIR methods, display very similar MIR
and X-ray surface brightness distributions.  We note however, that
there is tentative evidence that such X-ray bright clusters detected
using optical-MIR methods display numerically larger populations of
red sequence galaxies than X-ray selected counterparts, an observation
consistent with SpARCS selection based upon the presence of an
identifiable red sequence.

There are further physical differences between cluster samples
detected using each method. Within the optical-MIR selected SpARCS
sample we have identified three sub-samples of clusters based upon
their X-ray and MIR aperture flux measurements.  In particular, the
sub-sample of X-ray faint, MIR bright SpARCS clusters display the same
redshift distribution as X-ray bright SpARCS clusters yet show,
on-average, higher values of the BCG-barycentre distance.  In the
literature, larger values of BGC-cluster centroid offset have been
demonstrated to be strongly correlated with more shallow X-ray surface
brightness profiles \citep[e.g.][]{sanderson2009,mantz2015}.

Interpreting greater values of the BCG-centroid distance as an
indicator of dynamical disturbance in a cluster might provide an
explanation of their relatively low X-ray compared to stellar
emission, e.g. if the X-ray emitting gas has been disturbed by a
recent merger, reducing the X-ray surface brightness as a result
\citep[e.g.][]{eckert2011,barnes2017}.  An alternative explanation is
that the X-ray faint, MIR bright cluster sub-sample is associated with
secular mass assembly in massive structures whereby a compact,
low-mass, virialised core is surrounded by an envelope of accreting
material (c.f. Figure 1 of \citealt{muldrew2015}).  The location of
the BCG can be perturbed but, overall, the mass accretion is on-going
and continuous as opposed to associated with large, stochastic merger
events. Note that this statement regarding the dynamical state of optical-MIR
selected clusters does not ignore the observation that X-ray selected
clusters also display a range of relaxation states
\citep[e.g.][]{sanderson2009,lavoie2016}. However, the departures from
relaxation are slight compared to that observed in the present SpARCS
sample.

A contrasting argument, that such X-ray faint, MIR bright systems
represent projected large-scale structure variations along the
line-of-sight, as opposed to bound systems displaying incomplete
virialisation, requires a rate of contamination in marked disagreement
with previously determined rates of false detection in such systems
\citep{gladders2000} and with the successful results of spectroscopic
follow-up campaigns employing these clusters
\citep[e.g.][]{muzzin2012}. Of the $z>0.8$ SpARCS clusters with IRAC1
aperture fluxes $> 650 \mu$Jy, 8/10 are X-ray faint. The corresponding
value for clusters with aperture luminosities $L_K>2.5 \times 10^{13}
\rm L_\odot$ is 8/16.
 
The analysis presented in this paper shares many similarities with
that of \cite{rossetti2017} who compare clusters selected from the
Planck SZ catalogue with X-ray selected clusters from the MAssive
Cluster Survey \citep[MACS;][]{ebeling2010}. \cite{rossetti2017}
demonstrate that the X-ray cool-core fraction of SZ detected clusters
is significantly lower than that determined for X-ray selected
clusters and claim that this result can be explained in large part as
due to the relative detectability of clusters of varying surface
brightness properties in each sample. Interestingly, they also detect
a population of shallow surface brightness profile (NCC) SZ ``bright''
clusters that are undetected in MACS yet possess X-ray luminosities
based upon an extrapolated $L_X - Y$ relation that nominally place
them within the MACS selection criteria \citep[Figure 9
  of][]{rossetti2017}, i.e. X-ray under luminous for their
SZ-determined mass.

X-ray emission in the X-ray faint, MIR bright sub-sample of SpARCS
clusters is clearly present and associated with the location of the
BCG in each cluster. Figures \ref{xsb_flux_all} and \ref{xsb_lum_all}
indicate that the stacked X-ray surface brightness distribution in
these clusters is similar in shape to X-ray bright clusters yet offset
to lower overall normalisation.  This would appear to support the
assertion that X-ray faint, MIR bright SpARCS clusters represent
bona-fide clusters where a low-mass, virialised core surrounded by an
extended, bound envelope of material.

This conclusion is also supported by the observation that all samples
of clusters considered in this paper display identifiable red sequence
galaxy populations, thus confirming that we are observing real galaxy
overdensities of common age and star formation history as opposed to
chance projections.  If low mass groups are indeed the sites of
pre-processing to create such red sequence populations
\citep[e.g.][]{li2009} then these structures, accreting onto X-ray
faint, MIR bright SpARCS clusters may be responsible for the high
values of MIR aperture fluxes and luminosities observed in these
systems.  In addition, there is tentative evidence that the red
sequence population of X-ray selected clusters is marginally smaller
than the red sequence in optical-MIR selected clusters of comparable
X-ray brightness \citep[c.f.][]{donahue2002}.  However, the precision
achievable in this current study does not permit more than a tentative
statement.

It should also be noted that all of the above conclusions are based
upon the average properties of sub-samples of clusters and thus any
information of the distribution of relaxation states of optical-MIR
selected clusters is not available.  What we can say however, based
upon the distribution of clusters in the X-ray versus MIR aperture
flux plane, is that each SpARCS sub-sample is not isolated from any
other; instead each is drawn from a continuous range of properties
formed by the overall SpARCS sample and each is identified using
sensible, yet essentially arbitrary, cuts.

Furthermore, if the X-ray faint, MIR bright sub-sample of clusters is
indeed associated with either ongoing mass assembly onto a virialised
core or disruption from a recent major merger, then the X-ray surface
brightness profiles of such clusters would be expected to change
significantly as they evolve to a more relaxed state. The surface
brightness profile determines the detectability of faint clusters and
therefore modelling of X-ray cluster surface brightness profiles (or
equivalently the astrophysics underlying the surface brightness
profile) is a key factor in computing an accurate cluster selection
function. This point extends beyond the inclusion of an explicit
surface brightness expression for clusters in the selection modelling
\citep[e.g.][]{pacaud2006} to a description that includes a
distribution of surface brightness properties informed by the
dynamical state of the cluster population.  Cluster mass assembly
state therefore represents an important source of astrophysical
uncertainty, particularly in the application of X-ray selected cluster
samples to cosmological analyses employing cluster number counts
\citep[e.g.][]{borgani2001b,mantz2008}.

If the dynamical state of galaxy clusters is indeed causing surface
brightness driven incompleteness in observations of galaxy clusters,
how should such observations be reconciled with the aim of using
galaxy clusters as a probe of the cosmological model?  For
cosmological applications, one typically compares the observed sample
to the true population as provided by either a numerical model or
simulations.

Focussing on the relationship between mass growth in galaxy clusters
and their observability as a function of wavelength, if the X-ray
under luminous structures identified in this paper are collapsing
filaments, would they pass applied friends-of-friends or spherical
overdensity criteria to qualify as a halo and therefore contribute to
the mass function \citep[e.g.][]{watson2013}? If yes, then X-ray
selection functions might require an additional incompleteness term
(e.g. a sub-population of X-ray dark haloes in the scaling relation
model). If no, then optical-IR samples might require an additional
contamination term (e.g. to represent the selection of bound yet
unvirialised structures).  If X-ray under luminous clusters instead
represent recent mergers that have not yet reached equilibrium, one
might consider whether deblending methods are consistent between
observations and simulations.  Would numerical simulations identify
one or two halos?  Would X-ray image analysis do so, for example, were
deeper data available?

A related question is whether this X-ray under luminous population is
represented within scaling relation models.  Do such systems represent
simply the low-end of the log-normal distribution derived from
analysis of X-ray samples, or are they instead a population so far
relatively undetected by X-ray selected cluster studies? If they are
undetected, this would support the idea that X-ray scaling relations
should be derived from cluster samples selected at other wavelengths
e.g. \cite{andreon2016}.

In closing, it is clear that any survey for galaxy clusters provides
only a partial view of the true population of virial structures above
a given mass threshold.  However, comparisons of cluster samples
compiled at multiple wavelengths, such as performed in this paper and
others, provide a means to reveal the nature and extent of any bias.
Ultimately, and possibly with recourse to simulated clusters samples
incorporating both cosmological and gas physics
\citep[e.g.][]{mccarthy2017,barnes2017}, the effects of such bias can
be corrected for and an impartial view obtained of the formation of
large scale structure and the evolution of galaxies therein.

\section*{Acknowledgments}

The authors would like to thank the anonymous referee for their
comments that resulted in many improvements in the paper.  The authors
further wish to thank Dr. Irene Pintos Castro for checking our
XMM-SpARCS matching results.  MERC acknowledges support by the
Bonn-Cologne Graduate School of Physics and Astronomy (BCGS) and the
German Aerospace Agency (DLR) with funds from the Ministry of Economy
and Technology (BMWi) through grant 50 OR 1608. G.W. acknowledges
financial support for this work from NSF grant AST-1517863 and from
NASA through programs GO-13306, GO- 13677, GO-13747 \&\ GO-13845/14327
from the Space Telescope Science Institute, which is operated by AURA,
Inc., under NASA contract NAS 5-26555, and grant number 80NSSC17K0019
issued through the Astrophysics Data Analysis Program (ADAP).

\bibliographystyle{mn2e}
\bibliography{references}{}

\bsp

\label{lastpage}

\end{document}